\documentstyle[twoside]{article}

\catcode`\@=11
\long\def\@makefntext#1{
\protect\noindent \hbox to 3.2pt {\hskip-.9pt
$^{{\eightrm\@thefnmark}}$\hfil}#1\hfill}		

\def\thefootnote{\fnsymbol{footnote}}
\def\@makefnmark{\hbox to 0pt{$^{\@thefnmark}$\hss}}	

\def\ps@myheadings{\let\@mkboth\@gobbletwo
\def\@oddhead{\hbox{}
\rightmark\hfil\eightrm\thepage}
\def\@oddfoot{}\def\@evenhead{\eightrm\thepage\hfil
\leftmark\hbox{}}\def\@evenfoot{}
\def\sectionmark##1{}\def\subsectionmark##1{}}

\oddsidemargin=\evensidemargin
\renewcommand{\thefootnote}{\fnsymbol{footnote}}

\newcounter{sectionc}
\newcounter{subsectionc}
\newcounter{subsubsectionc}
\renewcommand{\section}[1] {\vspace{12pt}\addtocounter{sectionc}{1}
\setcounter{subsectionc}{0}\setcounter{subsubsectionc}{0}\noindent
	{\tenbf\thesectionc. #1}\par\vspace{5pt}}
\renewcommand{\subsection}[1] {\vspace{12pt}
\addtocounter{subsectionc}{1}\setcounter{subsubsectionc}{0}\noindent
	{\bf\thesectionc.\thesubsectionc.
        {\kern1pt \bfit #1}}\par\vspace{5pt}}
\renewcommand{\subsubsection}[1] {\vspace{12pt}
\addtocounter{subsubsectionc}{1}\noindent
        {\tenrm\thesectionc.\thesubsectionc.\thesubsubsectionc.
	{\kern1pt \tenit #1}}\par\vspace{5pt}}

\newcounter{appendixc}
\newcounter{subappendixc}[appendixc]
\newcounter{subsubappendixc}[subappendixc]
\renewcommand{\thesubappendixc}{\Alph{appendixc}.
        \arabic{subappendixc}}
\renewcommand{\thesubsubappendixc}{\Alph{appendixc}.
        \arabic{subappendixc}.\arabic{subsubappendixc}}

\renewcommand{\appendix}[1] {\vspace{12pt}
        \refstepcounter{appendixc}
        \setcounter{figure}{0}
        \setcounter{table}{0}
        \setcounter{lemma}{0}
        \setcounter{theorem}{0}
        \setcounter{corollary}{0}
        \setcounter{definition}{0}
        \setcounter{equation}{0}
        \renewcommand{\thefigure}{\Alph{appendixc}.\arabic{figure}}
        \renewcommand{\thetable}{\Alph{appendixc}.\arabic{table}}
        \renewcommand{\theappendixc}{\Alph{appendixc}}
        \renewcommand{\thelemma}{\Alph{appendixc}.\arabic{lemma}}
        \renewcommand{\thetheorem}{\Alph{appendixc}.\arabic{theorem}}
        \renewcommand{\thedefinition}{\Alph{appendixc}.
         \arabic{definition}}
        \renewcommand{\thecorollary}{\Alph{appendixc}.
         \arabic{corollary}}
        \renewcommand{\theequation}{\Alph{appendixc}.
         \arabic{equation}}
        \noindent{\tenbf Appendix \theappendixc #1}\par\vspace{5pt}}
\newcommand{\subappendix}[1] {\vspace{12pt}
        \refstepcounter{subappendixc}
        \noindent{\bf Appendix \thesubappendixc. {\kern1pt \bfit #1}}
	\par\vspace{5pt}}
\newcommand{\subsubappendix}[1] {\vspace{12pt}
        \refstepcounter{subsubappendixc}
        \noindent{\rm Appendix \thesubsubappendixc.
        {\kern1pt \tenit #1}}\par\vspace{5pt}}

\newcommand{\textlineskip}{\baselineskip=13pt}
\newcommand{\smalllineskip}{\baselineskip=10pt}

\def\eightcirc{
\begin{picture}(0,0)
\put(4.4,1.8){\circle{6.5}}
\end{picture}}
\def\eightcopyright{\eightcirc\kern2.7pt\hbox{\eightrm c}}

\def\abstracts#1#2#3{{
	\centering{\begin{minipage}{4.5in}\baselineskip=10pt
        \footnotesize
	\parindent=0pt #1\par
	\parindent=15pt #2\par
	\parindent=15pt #3
	\end{minipage}}\par}}

\renewenvironment{thebibliography}[1]
	{\frenchspacing
	 \ninerm\baselineskip=11pt
	 \begin{list}{\arabic{enumi}.}
	{\usecounter{enumi}\setlength{\parsep}{0pt}
	 \setlength{\leftmargin 12.7pt}{\rightmargin 0pt}
	 \setlength{\itemsep}{0pt} \settowidth
	{\labelwidth}{#1.}\sloppy}}{\end{list}}

\newcounter{itemlistc}
\newcounter{romanlistc}
\newcounter{alphlistc}
\newcounter{arabiclistc}

\newcommand{\fcaption}[1]{
        \refstepcounter{figure}
        \setbox\@tempboxa = \hbox{\footnotesize Fig.~\thefigure. #1}
        \ifdim \wd\@tempboxa > 5in
           {\begin{center}
        \parbox{5in}{\footnotesize\smalllineskip Fig.~\thefigure. #1}
            \end{center}}
        \else
             {\begin{center}
             {\footnotesize Fig.~\thefigure. #1}
              \end{center}}
        \fi}

\newcommand{\tcaption}[1]{
        \refstepcounter{table}
        \setbox\@tempboxa = \hbox{\footnotesize Table~\thetable. #1}
        \ifdim \wd\@tempboxa > 5in
           {\begin{center}
        \parbox{5in}{\footnotesize\smalllineskip Table~\thetable. #1}
            \end{center}}
        \else
             {\begin{center}
             {\footnotesize Table~\thetable. #1}
              \end{center}}
        \fi}

\def\@citex[#1]#2{\if@filesw\immediate\write\@auxout
	{\string\citation{#2}}\fi
\def\@citea{}\@cite{\@for\@citeb:=#2\do
	{\@citea\def\@citea{,}\@ifundefined
	{b@\@citeb}{{\bf ?}\@warning
	{Citation `\@citeb' on page \thepage \space undefined}}
	{\csname b@\@citeb\endcsname}}}{#1}}

\newif\if@cghi
\def\cite{\@cghitrue\@ifnextchar [{\@tempswatrue
	\@citex}{\@tempswafalse\@citex[]}}
\def\citelow{\@cghifalse\@ifnextchar [{\@tempswatrue
	\@citex}{\@tempswafalse\@citex[]}}
\def\@cite#1#2{{$\null^{#1}$\if@tempswa\typeout
	{IJCGA warning: optional citation argument
	ignored: `#2'} \fi}}

\def\pmb#1{\setbox0=\hbox{#1}
	\kern-.025em\copy0\kern-\wd0
	\kern.05em\copy0\kern-\wd0
	\kern-.025em\raise.0433em\box0}

\def\fnt#1#2{\footnotetext{\kern-.3em
	{$^{\mbox{\scriptsize #1}}$}{#2}}}

\def\fpage#1{\begingroup
\voffset=.3in
\thispagestyle{empty}\begin{table}[b]\centerline{\footnotesize #1}
	\end{table}\endgroup}

\font\tenrm=cmr10
\font\tenit=cmti10
\font\tenbf=cmbx10
\font\bfit=cmbxti10 at 10pt
\font\ninerm=cmr9

\font\eightrm=cmr8

\def\qed{\hbox{${\vcenter{\vbox{		
   \hrule height 0.4pt\hbox{\vrule width 0.4pt height 6pt
   \kern5pt\vrule width 0.4pt}\hrule height 0.4pt}}}$}}

\renewcommand{\thefootnote}{\fnsymbol{footnote}}

\input epsf
\global\arraycolsep=2pt
\def\spose#1{\hbox to 0pt{#1\hss}}
\def\lsim{\mathrel{\spose{\lower 3pt\hbox{$\mathchar"218$}}
 \raise 2.0pt\hbox{$\mathchar"13C$}}}
\def\gsim{\mathrel{\spose{\lower 3pt\hbox{$\mathchar"218$}}
 \raise 2.0pt\hbox{$\mathchar"13E$}}}

\renewcommand{\theequation}{\thesection.\arabic{equation}}
\def\laq{\raise 0.4ex\hbox{$<$}\kern -0.8em\lower 0.62
ex\hbox{$\sim$}}
\def\gaq{\raise 0.4ex\hbox{$>$}\kern -0.7em\lower 0.62
ex\hbox{$\sim$}}

\def\beq{\begin{equation}}
\def\eeq{\end{equation}}
\def\bea{\begin{eqnarray}}
\def\eea{\end{eqnarray}}

\def \pa {\partial}
\def \ra {\rightarrow}

\def \ti {\tilde}
\def \la {\lambda}

\def \La {\Lambda}
\def \Da {\Delta}
\def \b {\beta}
\def \a {\alpha}
\def \ap {\alpha^{\prime}}

\def \ga {\gamma}

\def \da {\delta}
\def \ep {\epsilon}
\def \r {\rho}
\def \om {\omega}
\def \Om {\Omega}
\def \noi {\noindent}

\begin{document}

\begin{titlepage}

\begin{flushright}
CERN-TH/96-186\\
hep-th/9607146
\end{flushright}

\vspace{2 cm}

\begin{center}
\Large\bf Relic Gravitons from the Pre-Big Bang: \\
\Large\bf What we Know and What we Do Not Know
\footnote{Based on talks presented at the Meeting on
{\sl ``Detection of high-frequency gravitational waves"} (CERN,
January 1996), at the INTAS Meeting on {\sl ``Fundamental
problems in classical, quantum and string gravity"} (Turin, May
1996), at the {\sl ``XLth Conference of the Italian Astronomical
Society"} (Osservatorio di Roma, Rome, May 1996), and at the
Network Meeting on {\sl ``String Gravity"} (Observatoire de
Paris, June 1996).}
\end{center}

\vspace{1.5cm}

\begin{center}
M. Gasperini\\
{\sl Theory Division, CERN, CH-1211 Geneva 23, Switzerland}\\
and\\
{\sl Dipartimento di Fisica Teorica, Universit\`a di Torino,}\\
{\sl Via P. Giuria 1, 10125 Turin, Italy}
\end{center}

\vspace{1.5cm}

\begin{abstract}
\noi
I discuss the status of present knowledge about a possible
background of relic gravitons left by an early, pre-big bang
cosmological epoch, whose existence in the past of our Universe
is suggested by the duality symmetries of string theory.
\end{abstract}

\vspace{1.5cm}
\begin{center}
To appear in \\
{\sl ``New developments in string gravity
and physics at the Planck energy scale"}\\
ed. by N. Sanchez (World Scientific, Singapore, 1996)
\end{center}
 \vspace{1.5cm}
\vfill
\begin{flushleft}
CERN-TH/96-186\\
July 1996 
\end{flushleft}

\end{titlepage}

\thispagestyle{empty}
\vbox{}
\newpage

\normalsize\textlineskip
\thispagestyle{empty}
\setcounter{page}{1}


\vspace*{0.88truein}

\fpage{1}

\centerline{\bf RELIC GRAVITONS FROM THE PRE-BIG BANG:}
\vspace*{0.15truein}
\centerline{\bf WHAT WE KNOW AND WHAT WE DO NOT KNOW}
\vspace*{0.37truein}

\centerline{\footnotesize MAURIZIO GASPERINI}
\vspace*{0.015truein}
\centerline{\footnotesize\it Theory Division, CERN,
 CH-1211 Geneva 23, Switzerland}
\baselineskip=10pt
\centerline{\footnotesize and {\it Dipartimento di Fisica Teorica, 
Universit\`a di Torino, Turin, Italy}}
\vspace*{0.225truein}

\vspace*{0.21truein}
\abstracts{I discuss the status of present knowledge about
 a possible
background of relic gravitons left by an early, pre-big bang
cosmological epoch, whose existence in the past of our Universe
is suggested by the duality symmetries of string theory.}{}{}
\vspace*{1pt}\textlineskip

\textheight=7.8truein
\setcounter{footnote}{0}
\renewcommand{\thefootnote}{\alph{footnote}}

\vspace*{0.3truein}
\centerline{\bf Table of Contents}
\vspace*{0.2truein}
\noi
{1. Introduction}

\noi
{2. The Pre-Big Bang Scenario}

{2.1. Motivations}

{2.2. Kinematical aspects}

\noi
{3. Growth of Perturbations}

{3.1. Formal problems}

{3.2. Phenomenological consequences}

\noi
{4. The Expected Graviton Background}

{4.1. Minimal model}

{4.2. A diagrammatic approach to the perturbation spectrum}

{4.3. Constraints from photon production}

{4.4. Non-minimal models}

{4.5. Experimental sensitivities}

\noi
{5. Conclusion}

{Appendix A: The dimensionless amplitude $|\da_h|$}

{Appendix B: The spectral amplitude $S_h$}

\newpage
\section{Introduction}
\noindent
The cosmological solutions of the low energy string effective
action, and their symmetry properties, have recently motivated 
the study 
of a very early, ``pre-big bang" cosmological phase\cite{1,2,3},
whose kinematical properties are the ``dual" counterpart of the
present, standard cosmological phase. The purpose of this paper
is to give a short introduction to the pre-big bang scenario, to
explain why the amplification of perturbations is more efficient
in such a context than in the standard inflationary scenario, and
to discuss the general properties of a possible cosmic graviton
background associated to the transition from the pre- to the
post-big bang epoch. 

The main ideas and results reported here are the fruit of a
research programme in string cosmology that started from earlier
work done with Norma Sanchez and Gabriele Veneziano on the
motion of strings in accelerated backgrounds\cite{4,5}, and that
was subsequently developed in collaboration with Gabriele
Veneziano and with a number of other researchers: Ramy 
Brustein, Massimo Giovannini, Jnan Maharana, Krzysz 
Meissner and
Slava Mukhanov (listed here in alphabetical order). The
interested reader can find an updated list of references and 
papers about the pre-big bang scenario on the WWW ``home
page" devoted to string cosmology, available at the address:

\centerline{\tt http://www.to.infn.it/teorici/gasperini/}
\vskip 1 cm

\renewcommand{\theequation}{2.\arabic{equation}}
\setcounter{equation}{0}
\section{The Pre-Big Bang Scenario}
\label{sec:2}
\noindent
I will call, for short, pre-big bang scenario, a cosmological 
scenario that is based on the equations obtained from the string
effective action, and which includes an initial phase of
accelerated evolution and growing curvature\cite{1,2,3}. The
most revolutionary aspect of the class of models describing this
scenario is probably the fact that the initial state of the
Universe, instead of being hot and dense as in the standard
cosmological context, is the string perturbative vacuum, namely a state with
flat metric, vanishing gauge coupling ($g=e^{\phi/2}=0$,
$\phi=-\infty$), and no matter present (with the possible
exception of an incoherent, highly diluted gas of non-interacting
strings\cite{3}). As a consequence, in this class of models the initial
evolution of the Universe can be consistently described in terms
of the lowest order string effective action,
\bea
S=&-&\int d^{d+1}x\sqrt{|g|}\left[e^{-\phi}\left(R+(\nabla
\phi)^2+ \ap{\rm corrections}\right) +V(\phi, {\rm
nonperturbative})\right] \nonumber\\
&+&{\rm loops}\left(e^\phi\right) ,
\label{21}
\eea
neglecting both finite-size effects ($\ap$ corrections) and higher
(field theory) loops in the  coupling constant $g=e^{\phi/2}$.
The initial evolution is simply driven by the kinetic energy of the
dilaton field, with negligible contributions also from a possible
non-perturbative potential $V(\phi)$, which is known, at small
coupling, to approach zero very rapidly\cite{6} as $V\sim
\exp[-\exp (-\phi)]$. Here 
I will neglect, for simplicity, also a possible
anisotropic contribution of the string antisymmetric tensor to
the cosmological equations; see Ref. [7] for the inclusion of such
a contribution. 

Irrespective of the possible initial presence of 
matter sources\cite{2,3}, 
the (generally anisotropic) solution of the string
cosmology equations, after inserting the perturbative vacuum as
initial condition, can be asymptotically parametrized as
\beq
a_i=(-t)^{\b_i}, ~~~ \phi= \left(\sum_i \b_i-1\right)\ln(-t), ~~~
\sum_i\b_i^2=1, ~~~ t<0, ~~~ t\ra 0
\label{22}
\eeq
($i=1,..,d$). This solution describes a phase in which the curvature scale
and the dilaton are both growing, since $|H_i|\sim
(-t)^{-1} \sim |\dot \phi|$ for $t\ra 0_-$, where $H=\dot a/a$, and
the dot denotes derivatives with respect to cosmic time $t$. 
A typical example of background is illustrated in Fig. 1. 

\begin{figure}[htb]
   \epsfxsize=6cm
   \centerline{\epsfbox{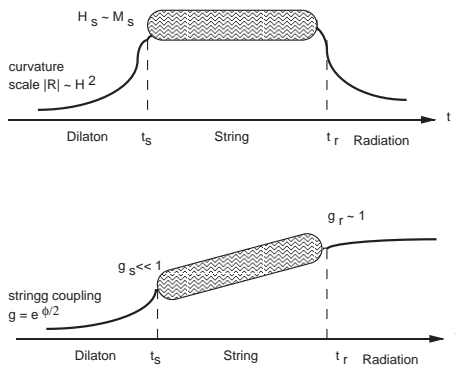}}
   \centerline{\parbox{11.5cm}{\caption{\label{fig:f1}
Qualitative evolution of the curvature scale and of the string
coupling as a function of time, for a generic pre-big bang type of
background. }}}
\end{figure}

Of course, when the curvature reaches the string scale,
determined by the string mass parameter $M_s\equiv
\la_s^{-1}$, the background enters a truly ``stringy" regime, in
which all higher derivative terms in the $\ap$ expansion become
important (our present imperfect knowledge of the detailed evolution, 
in that region, is represented by the wavy lines  
of Fig. 1). At the end of the string phase the dilaton
eventually approaches the strong coupling regime,
$g_r\equiv g(t_r)\sim 1$, where both the quantum loop effects 
and the 
semiclassical back-reaction of the produced radiation become
important, and the transition to the standard decelerated evolution
(with frozen dilaton) is expected to  occur. Some
concrete examples of how the inclusion of one-loop corrections
can stop the growth of the curvature\cite{6a} have been 
presented  recently and discussed in the context of four-\cite{7a} and
two-dimensional\cite{7b} cosmology. 

At the beginning of the string phase, on the contrary, the value
of the string coupling $g_s\equiv g(t_s)$ can be arbitrarily small
(see Fig. 1); it represents one of the main parameters of this
class of backgrounds. The other important parameter is the
duration of the string phase, measured by the ratio $|t_s/t_r|$, 
or by the ratio of the associated scale factors,
$a(t_s)/a(t_r)$. If $g_s$ is already of order $1$ at the beginning
of the string phase, the duration of such a phase presumably
shrinks to a number of order unity (in string units). The
presence of such a phase, however, cannot be avoided;  
otherwise, to the tree-level in $g_s$, and to zeroth order in
$\ap$, the growth of the curvature and of the dilaton is
unbounded\cite{8}. Even in that case, however, one can
compute quantum-mechanically, via the Wheeler-De Witt
equation, the transition probability from the initial to the final
background configuration. Such a transition can be represented
as a wave reflection in minisuperspace\cite{9,10}, and the
probability turns out to be finite and non-vanishing even in the
presence of a singularity that disconnects, classically, the two
asymptotic regimes.

\subsection{Motivations}
\noindent
There are at least two motivations, in a string theory context,
that lead to introduce an early, pre-big bang cosmological phase
in the past history of our Universe, as a natural complement of
the present, post-big bang cosmological era. The first 
motivation is provided
by the solutions of the classical string equations of motion
embedded in inflationary cosmological backgrounds\cite{4,5}. In 
the presence of shrinking event horizons, a gas of classical string
develops in fact an effective negative pressure, so that the background
evolution can be consistently self-sustained by the string gas
itself as a source\cite{3,11}. 

The second motivation is provided by the duality symmetries of
the lowest-order string effective action, which can be written, 
including the
antisymmetric tensor $B_{\mu\nu}$,
\beq
S=-\int d^{d+1}x\sqrt{|g|}e^{-\phi}\left[R+\left(\nabla_\mu 
\phi\right)^2-{1\over
2}\left(\pa_{[\mu}B_{\nu\a]}\right)^2\right], ~~~\mu,\nu= 1,2,...,
d+1. 
\label{23}
\eeq
If $a(t)$ and $\phi(t)$ represent an exact, homogeneous and
isotropic solution of this action (with $B=0$ and zero spatial
curvature), then
\beq
\ti a=a^{-1}, ~~~~~~ \ti\phi= \phi -2 d\ln a
\label{24}
\eeq
represent another exact solution, thanks to the scale factor
duality symmetry\cite{12,13}. This symmetry is only a particular
case of a more general covariance property of the action
(\ref{23}) under global $O(d,d)$ transformations\cite{14,15},
which mixes non-trivially the spatial components of the metric
and of the antisymmetric tensor,
\beq
M \ra \ti M= \La^TM\La , ~~~~
M=\pmatrix{G^{-1} & -G^{-1}B \cr
B G^{-1} & G-BG^{-1}B\cr}
\label{25}
\eeq
($\La \in O(d,d)$ and $G=g_{ij}$, $B=B_{ij}$). Such an $O(d,d)$
covariance is also preserved by the addition of source terms to
the cosmological equations\cite{16}, provided they represent
``bulk" string matter, whose components satisfy the string
equation of motion in the given background. In the presence of
sources, and in the perfect fluid approximation, a scale factor
duality transformation is simply associated to a ``reflection" of
the equation of state\cite{12}, $p/\r \ra -p/\r$. 

The importance of these symmetries\cite{1}, in our cosmological
context, follows from the fact that by operating simultaneously 
a duality
transformation (for instance according to eq. (\ref{24})) and 
a time-reversal transformation on a given decelerated, post-big
bang solution, 
\beq
\dot a>0, ~~~~~\ddot a <0, ~~~~~ \dot H <0,
\label{26}
\eeq
it is always possible to obtain a new accelerated solution, with
growing curvature, of the pre-big bang type: 
\beq
\dot a>0, ~~~~~\ddot a >0, ~~~~~ \dot H >0.
\label{27}
\eeq
In $d=3$, for instance, the standard radiation-dominated
solution 
\beq
a=t^{1/2}, ~~~~~ \phi= {\rm const}, ~~~~~p=\r/3,  
~~~~~ t>0 
\label{28}
\eeq
(which is still an exact solution for the action (\ref{23})
supplemented by perfect fluid sources) is mapped to the pre-big
bang solution
\beq
a=(-t)^{-1/2}, ~~~~~ \phi= -3 \ln (-t), ~~~~~p=-\r/3, 
~~~~~ t<0
\label{29}
\eeq
(see also Refs. [18,19] for other, less trivial examples).

This natural association of pre- and post-big bang solutions,
which always come in pair from the lowest-order string
cosmology equations, is impossible in the standard cosmological
context based on the Einstein equations: in that case there is in
fact no field playing the role of the dilaton, and the above
duality symmetries cannot be implemented. The challenge is, of
course, to formulate a complete and consistent string
cosmology scenario in which the two duality-related solutions
are smoothly joined, in the context of a background satisfying
(probably only asymptotically) the self-dual condition\cite{1}
$a(t)=a^{-1}(-t)$. We shall assume in the rest of the paper that
such a smooth transition (impossible to lowest order) can be
successfully implemented when higher-order corrections are
included, and we shall discuss some phenomenological aspects 
of the emerging cosmological scenario.

\subsection{Kinematical aspects}
\noindent
From a kinematical point of view, the epoch of pre-big bang
evolution can be invariantly characterized as a phase of
shrinking event horizons\cite{1}. This means that the scale
factor $a(t)$, for an isotropic phase of pre-big bang evolution,
can be parametrized in cosmic time as
\beq
a(t)=(-t)^\b, ~~~~~~\b<1, ~~~~~~ t<0,
\label {10}
\eeq
where the condition $\b<1$ is required for the convergence of
the integral defining the comoving size of the event horizon:
\beq
\int_t^0{dt'\over a(t')} <\infty  ~~~ \Rightarrow ~~~
\b<1.
\label{211}
\eeq

\noi
This condition can be satisfied in two ways:

\begin{itemize}
\item{}$\b<0 \Rightarrow \dot a>0, \ddot a>0, \dot H>0$,\\
which corresponds to a metric describing a phase of accelerated
expansion and growing curvature, usually called
superinflation\cite{18} (or pole inflation).
\item{}$0<\b<1 \Rightarrow \dot a<0, \ddot a<0, \dot H<0$,\\
which corresponds to a metric describing accelerated
contraction and, again, growing curvature scale.
\end{itemize}

\noi
The first type of metric provides a representation of the pre-big
bang scenario in the String frame (also called Brans-Dicke
frame), in which test strings move along geodesic surfaces, the
second in the more conventional Einstein frame, in which the
gravi-dilaton action appears diagonalized in the standard,
canonical way. The String frame was explicitly used above for
presenting the effective action (\ref{21}) and the duality
transformations, the Einstein frame will be used below for
discussing the amplification of perturbations. The choice of the
frame is indeed a matter of convenience, as both frames give
consistent and physically equivalent descriptions of the pre-big
bang scenario (see Refs. [2,3] for the explicit relation between
the two frames and for a discussion of their physical
equivalence). 

In the context of this paper, what is important is the fact that 
both classes of backgrounds, 
irrespective of the sign of $\b$,  are accelerated, and in both 
classes the amplification of perturbations is
more efficient, at high frequency, than in a standard
inflationary background. This occurs for two reasons:

\begin{itemize}
\item{}First because the curvature scale is growing and, as a
consequence, the perturbation spectrum\cite{19,20} grows with
frequency\cite{18,20a};
\item{}Second because the comoving amplitude of perturbations
may grow in time even outside the horizon\cite{2,20b}, instead
of being frozen as in the standard scenario.
\end{itemize}

\noi
These two effects are crucial for obtaining a strong graviton
background at high frequency, and will thus be discussed in
some detail in the next section.

\vskip 1 cm

\renewcommand{\theequation}{3.\arabic{equation}}
\setcounter{equation}{0}
\section{Growth of perturbations}
\label{sec:3}
\noindent
For an accelerated background, the scale factor can be
conveniently parametrized in conformal time ($\eta$, such that
$dt =a d\eta$), and in the negative time range, as 
\beq
a(\eta)= (-\eta)^\a, ~~~~~~~~~~\eta <0. 
\label{30}
\eeq
Positive $\a$ corresponds to accelerated contraction, negative
$\a$ to accelerated expansion, $\a=-1$ corresponds in
particular to the standard de Sitter-like inflation (see Fig. 2). 

\begin{figure}[htb]
   \epsfxsize=7cm
   \centerline{\epsfbox{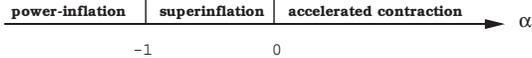}}
   \centerline{\parbox{11.5cm}{\caption{\label{fig:2}
Different classes of accelerated backgrounds, parametrized in
conformal time by the power $\a$.   }}} 
\end{figure}

Consider for instance the canonical variable $h_{\mu\nu}$
representing tensor perturbations, in the Einstein frame,
normalized to an initial vacuum fluctuation spectrum, i.e. 
$h_k\sim \left(e^{-ik\eta}/aM_p\sqrt k\right)$, for $\eta\ra
-\infty$ ($M_p$ is the Planck mass). 
The Fourier component of each polarization mode
$h_k$ satisfies the well-known equation\cite{19,20}
\beq
\left(ah_k\right)'' +\left(k^2-{a''\over a}\right) a h_k=0 . 
\label{31}
\eeq
In a background of the pre-big bang type the horizon is
shrinking, so that all modes are ``pushed out" of the horizon, for
$\eta\ra 0_-$.  For a mode whose wavelength is larger than the
horizon size ($|k\eta|<<1$), the asymptotic solution of 
the above perturbation equation 
is given by
\beq
h_k= A_k +B_k \left|\eta\right|^{1-2\a}, ~~~~~ \eta \ra 0_-
\label{32}
\eeq
where $A_k$ and $B_k$ are integration constants. We have in
general two possibilities.

\begin{itemize}
\item{}If $\a<1/2$ the first term is dominant in the asymptotic
solution  (\ref{32}), the comoving amplitude $h_k$ tends to stay
constant outside the horizon, and the typical (dimensionless)
perturbation amplitude $\da_h(k)$ (see Appendix A) can be
expressed as usual in terms of the Hubble factor at horizon
crossing: 
\beq
\left|\da_h(k)\right|\equiv k^{3/2}\left|h_k\right|\simeq
\left(H\over M_p\right)_{h.c.} , ~~~~~
{\rm horizon ~~crossing} \Leftrightarrow \left|k\eta\right|
\sim1. 
\label{33}
\eeq
This amplitude is constant in time. According to the horizon
crossing condition, however, higher-frequency modes cross the
horizon later in time, and then at higher value of the Hubble
factor if the curvature scale is growing, as in our class of
backgrounds. As a consequence, the amplitude of 
higher-frequency modes is enhanced with respect to 
lower frequency
modes, in contrast with the usual scale-invariant
(Harrison-Zeldovich) spectrum, where the amplitude is the same
for all modes.

\item{}If $\a>1/2$ the solution (\ref{32}) is dominated by the
second term, and $h_k$ tends to grow outside the horizon. The
typical perturbation amplitude becomes time-dependent,
\beq
\left|\da_h(k, \eta)\right|\simeq \left(H\over M_p\right)_{h.c.}
\left|k\eta\right|^{1-2\a}, ~~~~~\eta \ra 0_-,
\label{34}
\eeq
and grows not only with frequency, but also with time. This second
effect is not automatic, however, in any background of the
pre-big bang type, but requires a fast enough contraction (in the
Einstein frame) of the scale factor.
\end{itemize}

\subsection{Formal problems}
\noindent
The analysis of the perturbation spectrum, in the context of the
pre-big bang scenario, is complicated in general by two types of
formal problems. 

The first is that, just because of this ``anomalous" growth in
time of the amplitude, the linear approximation tends to break
down for scalar perturbations, in the standard longitudinal
gauge. This problem may be solved, at least for a class of
dilaton-driven backgrounds, by moving to an off-diagonal
gauge\cite{21} (also called ``uniform-curvature"
gauge\cite{21a}), in which the scalar part of the perturbed
metric is parametrized by two scalar potentials, $\Phi$ and $B$,
as follows
\beq
ds^2=a^2\left[d\eta^2(1+2\Phi)-dx_i^2-2\pa_i B dx^i d\eta
\right] . 
\label{35}
\eeq
In this way the dangerous  growing mode is ``gauged down",
enough to restore the validity of the linear approximation.  For
other backgrounds, however, the growth of perturbations
outside the horizon may be a true physical effect, which cannot be
eliminated in a convenient gauge; we have to restrict 
ourselves to a 
reduced portion of parameter space for the linear approximation
to be valid. 

A second problem 
is related to the normalization of the perturbations
to the initial spectrum of the quantum fluctuations of the
vacuum, in a background that is in general higher-dimensional
and anisotropic. Such a normalization requires the knowledge of
the so-called ``normal modes" of oscillation of an anisotropic
background, i.e. of the variables that diagonalize the action
perturbed to second order, and that satisfy canonical commutation
relations. This problem has been solved, up to now, only for
spatially flat higher-dimensional backgrounds, in which the
translations along internal dimensions are isometries of the full
perturbed metric\cite{22}. 

In any case, even if we have to restrict ourselves 
to a limited class of 
backgrounds in which the linear approximation is valid, and the
canonical normalization is known, we find that the enhanced
amplification of perturbation, in the context of the pre-big bang
scenario, is in general associated to new and interesting
phenomenological consequences.

\subsection{Phenomenological consequences}
\noindent
There are two effects, in particular, worth mentioning. The first
is the amplification of the vacuum fluctuations of the
electromagnetic field (and of the gauge fields, in general), due to
their coupling to a dynamical dilaton, according to the effective
interaction
\beq
e^{-\phi}\sqrt{-g}F_{\mu\nu}F^{\mu\nu}. 
\label{36}
\eeq

The electromagnetic field is also coupled to the metric, but in
four dimensions such a coupling is conformally invariant, and
there is no parametric amplification of perturbations in a
conformally flat metric background\cite{23}, like that of a typical
inflationary model. In a string cosmology context, the
amplification of the electromagnetic perturbations is a
consequence of the dilaton running during the phase of pre-big
bang evolution. Such an amplification may be large enough to
produce the ``seed fields" required for the origin of the galactic
magnetic field\cite{24}, and possibly also to produce the
observed anisotropy\cite{25} of the Cosmic Microwave
Background (CMB) radiation\cite{26}. 

A second important effect is the production of a relic graviton
background, much stronger, at high frequency, than predicted by
the standard inflationary scenario\cite{27}. 

The present spectral
energy density of such a background, in units of critical energy
density $\r_c$, is defined as
\beq
\Om_G(\om, t_0)={\om\over \r_c}{ d\r_G(\om, t_0)\over d\om},
~~~~~~~\r_c(t_0)={3M_p^2H_0^2\over 8\pi}, 
\label{37}
\eeq
where
\beq
{d\r_G(\om, t_0)\over d\om}=2\om\overline n (\om)
\left[4\pi\om^2\over (2\pi)^3\right],
\label{37a}
\eeq
and where $\overline n (\om)$ is the number density of produced
particles with proper energy $\om(t)=k/a(t)$ (the factor 2 is
because of the two graviton polarization modes; the factor in
square brackets refers to integration in momentum space). The
peak value of $\Om_G$, for any model of cosmological evolution,
can be conveniently referred to the present CMB energy density
$\Om_\ga$ (see next section), which is a number of order\cite{27a} 
$10^{-4}$--$10^{-5}$ (the uncertainty depends on the present
uncertainty  about the 
value of $\r_c$). In the standard inflationary scenario the energy 
density of 
the graviton background is constrained by the high
degree of isotropy of the CMB radiation\cite{25}, 
$(\Da T/T)_\ga~ \laq ~10^{-5}$, 
which imposes, at all scales $\om > 10^{-16}$Hz, the
bound\cite{28,29}
\beq
\Om_G~\laq ~\Om_\ga\left(\Da T \over T\right)_\ga^2~\laq 
~10^{-15} . 
\label{38}
\eeq

In the context of the pre-big bang scenario, on the contrary, the
graviton background is in general too low, on the COBE scale
$\om_0 \sim 10^{-18}$Hz, to be constrained by the CMB
isotropy\cite{3,21,27}. The peak value of the spectrum is simply 
controlled by the present value of the fundamental ratio of the
string and Planck mass, which is expected\cite{30} to be a
number in the range $0.3$--$0.03$:  
\beq
\Om_G~ \laq ~ \Om_\ga\left(M_s \over M_p\right)^2~\laq 
~10^{-6} . 
\label{39}
\eeq
This implies a possible signal, in the frequency range of present 
and near-future gravity wave detectors, up to {\em nine} orders
of magnitude higher than predicted by the standard inflationary
scenario. The next section is devoted to a detailed discussion of
this possibility. 

\vskip 1 cm

\renewcommand{\theequation}{4.\arabic{equation}}
\setcounter{equation}{0}
\section{The Expected Graviton Background}
\label{sec:4}
\noindent

\subsection{Minimal model of pre-big bang}
\noi
In order to discuss the expected graviton background we shall
first consider a ``minimal" model of pre-big bang scenario (the
same as used in Refs. [29, 32]), in which there are only three
phases: an initial phase of dilaton-driven inflation for $t<t_s$, a
high curvature string phase for $t_s<t<t_r$, and the standard
radiation-dominated evolution for $t>t_r$ (see also Fig. 1). 

At low frequency, namely for modes with
$\om<<\om_s \simeq (a_s\eta_s)^{-1}$, which   
cross the horizon before the 
beginning of the string phase, the slope of the spectrum can be
computed exactly from the perturbation equation obtained from
the low-energy action, and it is found to be cubic\cite{3,21,27}
(modulo small  corrections due to a logarithmic growth in time
of the comoving  amplitude  outside the horizon,
during the dilaton-driven phase): 
\beq
\Om_G(\om,t_0)\propto \left(\om\over
\om_s\right)^3\ln^2\left(\om\over \om_s\right) , ~~~~~~
\om<\om_s . \label{41}
\eeq

In this range of frequencies, the spectrum can easily be 
computed by noting that, for the given model of background, the
tensor perturbation equation (\ref{31}) reduces to a
Schr\"odinger-like equation in conformal time\cite{19,20}: 
\beq
\psi_k'' +\left[k^2-V(\eta)\right]\psi_k=0,
\label{42}
\eeq
with an effective potential ($V=a''/a$ in the Einstein frame)
whose modulus grows during the dilaton phase, keeps growing
in the string phase where it reaches a maximum around the
transition scale $\eta_r$, and rapidly approaches zero at the
beginning of the radiation era (see Fig. 3). 

\begin{figure}[htb]
   \epsfxsize=6cm
   \centerline{\epsfbox{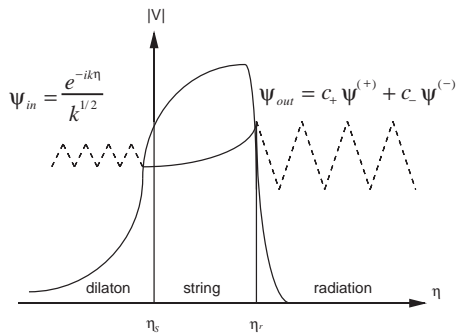}}
   \centerline{\parbox{11.5cm}{\caption{\label{fig:f3}
Parametric amplification of the perturbation mode $\psi_k$ due
to the effective potential barrier $V=a''/a$ (``anti-tunnelling" 
effect). }}}
\end{figure}

Normalizing the solution at $\eta \ra-\infty$ to an initial
vacuum fluctuation spectrum, $\psi_k\sim e^{-ik\eta}/\sqrt k$,
the asymptotic solution at late times in the radiation era, for
$\eta\ra +\infty$, is in general a linear combination of modes 
of positive and negative frequency with respect to
the initial vacuum state:
\beq
\psi_k=c_+\psi_k^{(+)}+c_- \psi_k^{(-)}, ~~~~~~
\eta \ra +\infty, 
\label{43}
\eeq
where the mixing coefficients $c_\pm$ parametrize the
unitary Bogoliubov transformation connecting the set of $|{\rm
in}\rangle$ and $|{\rm out}\rangle$ annihilation and creation
operators, 
\beq
\{\psi_{in}, b_k, b_k^\dagger\}  \Longrightarrow 
\{\psi_{out}, a_k, a_k^\dagger\},
\nonumber
\eeq
\beq
a_k=c_+b_k+c_-^\ast b_{-k}^\dagger, ~~~~~~
a_{-k}^\dagger=c_-b_k+c_+^\ast b_{-k}^\dagger .
\label{44}
\eeq
So, even starting from the vacuum, one ends up with a final
expectation number of gravitons, which is, in general,
non-vanishing\cite{30a,30b,30c}
\beq
\overline n_i=\langle0|b^\dagger b|0\rangle , ~~~~~
\overline n_f=\langle0|a^\dagger a|0\rangle =|c_-|^2 \not= 0 .
\label{45}
\eeq
The Bogoliubov coefficients $c_\pm(k)$ can be computed by
solving the perturbation equation (\ref{42}) in the three
different regions of Fig. 3, and by matching $\psi(\eta)$,
$\psi'(\eta)$ at $\eta=\eta_s$ and $\eta=\eta_r$. The
expectation number $\overline n(k)=|c_-(k)|^2$ then gives the
spectral energy distribution $\Om_G(\om)$ according to eqs.
(\ref{37}), (\ref{37a}).

This process of parametric amplification\cite{19} can be seen as
an ``anti-tunnelling" effect for the ``wave function" $\psi_k$.
Indeed, if we look at Fig. 3 from the right to the left, and
imagine to replace the time variable $\eta$ with a spacelike
coordinate, we obtain the description of a scattering process in
which $\psi^{(+)}\sim e^{-ik\eta}$ plays the role of a wave
incident on the barrier from $+\infty$, $\psi^{(-)}\sim
e^{+ik\eta}$ is the reflected wave, and $\psi_{in}$ is the
transmitted wave, 
\bea
\eta \ra +\infty &,& ~~~~~~~ \psi \sim A_{in} e^{-ik\eta}+
 A_{ref} e^{ik\eta}\nonumber \\
\eta \ra -\infty &,& ~~~~~~~ \psi \sim  A_{tr} e^{-ik\eta}.
\label{44a}
\eea
In the semiclassical approximation, which corresponds in our case
to the limit of a very large number of produced particles,
$\overline n_f >>1$, the reflection
coefficient $R=|A_{ref}|^2/|A_{in}|^2$ is approximately $1$, and
the Bogoliubov coefficient $|c_-|^2$ becomes the inverse of the
tunnelling coefficient $T=|A_{tr}|^2/|A_{in}|^2$:
\beq
|c_-|^2= {|A_{ref}|^2\over|A_{tr}|^2}={R\over T}\simeq {1\over T}.
\label{44b}
\eeq
Hence the name anti-tunnelling.

At high frequencies, namely for modes with $\om>>\om_s$,
crossing the horizon (or ``hitting" the effective potential
barrier) during the string phase, the exact slope of the spectrum
is at present unknown. In fact, even assuming that the
background curvature stays constant, $H\simeq M_s$ (in string
units) during the string phase\cite{24,27}, we do not know the
corresponding evolution of the dilaton background. Also, and
most important, we do not know the exact form of the
perturbation equation, because of possible higher derivatives
corrections due to higher orders in the $\ap$ expansion
(eventually, also higher-loop contributions cannot be excluded, a
priori).

It is important to stress, however, that the latter problem can
at most affect the amplitude, but not the slope (\ref{41}) of the
``dilatonic" branch of the spectrum. In fact, a transverse and
traceless mode of constant amplitude is always a solution of a
linearized and covariant perturbation equation, to all orders in
the background curvature\cite{31}. This preserves the spectral
distribution of modes whose amplitude is frozen already during
the dilaton-driven phase.

Concerning the former problem, we can parametrize our
ignorance of the background evolution with a power-like
behaviour of the dilaton background\cite{24,27} (see also next
section). By using the low-energy perturbation equation, we can
thus predict for the graviton spectrum a slope that is at most
cubic, but in general flatter than at lower frequencies.
Interestingly enough, this property is valid for the {\em exact }
solution of the low energy perturbation equation\cite{31}, to all
orders in the $k$ expansion (\ref{32}), and for any given
time-dependent background, not only of the power-law type.
Since this result is a consequence of the duality symmetry
characterizing the low-energy evolution of
perturbations\cite{31a}, one might argue that a flatter slope at
higher frequency is in general to be expected, provided the
duality symmetry is not spoiled by higher derivatives 
and higher-loop corrections. 

What we can do, also, irrespective of a detailed knowledge of
the background kinematics, is to provide an accurate estimate
of the present value of the end point of the spectrum, namely of
the maximal frequency undergoing parametric
amplification\cite{32}. This frequency is of the same order as the
frequency corresponding to the production of one graviton per
polarization and per unit phase-space volume (it is known that,
for higher frequencies, graviton production is exponentially
suppressed\cite{23,30b}). To this aim we rescale such a
frequency, which we call $\om_1(t)$, to the beginning of the
radiation era, i.e. $\om_1(t_0)=\om_1(t_r) a_r/a_0$, where by
definition
\beq
H_r^2={8\pi \over 3M_p^2}{\pi^2 N_r\over 30} T_r^4
\label{45}
\eeq
($N_r$ is the total effective number of massless
degrees of freedom in thermal equilibrium at $t=t_r$). Also, we
rescale $T_r$ to the present CMB temperature $T_0=2.7$K, by
taking into account a possible production of radiation due to
some reheating process occurring well below the string scale.
The effect of such processes can be parametrized by the
associate fraction of present thermal entropy density, $\da s$, as 
\beq
s_0\equiv {2\pi^2\over45}n_0(a_0T_0)^3= {2\pi^2\over45}
n_r(a_rT_r)^3+s_0\da s ,
\label{46}
\eeq
where $n_0$ and $n_r$ are the number of species contributing
(with their own statistical weight) to the thermal entropy at
$t_0$ and $t_r$, respectively ($n_0=3.91$, see for instance Ref.
[33]). 

In a string cosmology context, the amplification of tensor
perturbations is always accompanied by a copious production of
gauge bosons (according to eq. (\ref{36})) and other
ultrarelativistic particles, which are expected to thermalize
rapidly, unlike gravitons and dilatons, which interact only
gravitationally\cite{35,35a}. For the formulation of a consistent
scenario in which the energy density of such particles, produced
from the vacuum, becomes critical and start dominating the
background evolution at $t=t_r$, we shall assume that  both
$H_r$ and $\om_1(t_r)$ are of the same order as the string 
scale at $t_r$, i.e. $H_r\simeq M_s(t_r)\simeq \om_1(t_r)$. In
that case we obtain, from eqs. (\ref{45}) and (\ref{46}),
\beq
\om_1(t_0)\simeq 1.1~
 T_0\left[M_s(t_r)\over M_p\right]^{1/2}
\left(10^3\over n_r\right)^{1/12}
(1-\da s)^{1/3} ,
\label{47}
\eeq
where we have used the fact that $n_r\simeq N_r$, and 
that $N_r$ is expected
to be a number of order $10^2$--$10^3$. 

On the other hand, the fact that the radiation temperature
cannot exceed the string mass scale in a string cosmology
context, $T_r~\laq~ M_s(t_r)$, implies that the total energy
density of the quantum fluctuations (equal to $\r_r$, at $t=t_r$) 
is dominated by the energy density $\r_1$ corresponding to the
end-point frequency $\om_1$, i.e. $\r_1=\om_1^4/\pi^2$. 
In fact, 
\beq
{\r_r\over N_r\om_1^4(t_r)}\simeq
{\rho_r\over N_r M_s^4(t_r)}= {\pi^2  \over 30}{T_r^4\over 
M_s^4(t_r)}~\laq~ 1 .
\label{48}
\eeq
In addition, the fact that $\r_r\simeq N_rM_s^4$ is critical at 
$t=t_r$ implies that the string scale is already quite close to its
present value, as $M_p/M_s(t_r)\simeq N_r^{1/2}\sim 10$--$30$.
Consequently, the peak of the graviton background must be of
the same order as the end-point value $\Om_G(\om_1,
t_0)=\om_1(t_0)/\pi^2\r_c(t_0)$, with $M_s$ fixed by a dilaton
expectation value already in its present range. 

So, the spectral graviton
distribution has to be on the average non-decreasing from
$\om_s$ to $\om_1$, with an upper bound that can be written
in critical units as\cite{32}
\bea
\Om_G(\om_1,t_0)&\simeq& 2.6~\Om_\ga(t_0)\left(M_s\over
M_p\right)^2 \left(10^3\over n_r\right)^{1/3}\left(1-\da
s\right)^{4/3}\nonumber\\
&\simeq&7\times 10^{-5}h_{100}^{-2}\left(M_s\over
M_p\right)^2 \left(10^3\over n_r\right)^{1/3}\left(1-\da
s\right)^{4/3}, 
\label{49}
\eea
where $h_{100}=H_0/(100 ~{\rm km ~ sec}^{-1}{\rm Mpc}^{-1})$. 
Note that
this peak value can be integrated from $\om_1$ down to the
Hertz scale without contradicting the bound following from the
standard nucleosynthesis analysis\cite{27a},
\beq
h^2_{100}\int \Om_G (\om, t_0) d\ln \om~ \laq~ 0.5 \times 
10^{-5}. 
\label{410}
\eeq

The above situation is illustrated in Fig. 4, which shows the
maximum allowed graviton energy density for the case
$n_r=10^3$, in the Hz to GHz range. For any given $\da s$, the
uncertainty of the end point (\ref{47}) and of the peak (\ref{49})
is mainly due to the present theoretical uncertainty of the
fundamental ratio $M_s/M_p$, which has been chosen, 
for the illustrative purpose
of Fig. 4, to range from $0.1$ to $0.01$. 
This uncertainty is represented by the shaded boxes of Fig. 4. To the
left of the end point the spectrum can be at most flat, so that 
in the absence of significant reheating at scales much lower than
$t_r$, the expected maximal strength of the graviton
background should lie within the dashed lines in the band
labelled by $\da s=0$. 

A non-zero $\da s$ would instead imply
that the radiation that becomes dominant at the end of the
string phase is only a fraction of the thermal radiation that we
observe today, with a consequent dilution of the graviton
background (produced at the string scale) with respect to the
total present radiation energy density. However, we can see
from the picture that even if $99 \%$ of the present entropy
would be produced during the latest stage of evolution, the
intensity of the relic graviton background would stay well above the
full line labelled ``de Sitter" in Fig. 4, which represents the most
optimistic prediction of the standard inflationary
scenario\cite{29}, in this range of frequency. 

Also shown in the picture are three lines of constant sensitivity
of a gravity wave detector, $S_h^{1/2}=10^{-23}, 10^{-25}, 
10^{-27}$Hz$^{-1/2}$, given in terms of the spectral amplitude
$S_h^{1/2}$, related to the spectral energy density $\Om_G$ by
\beq
S_h(\nu)={3H_0^2\over 4\pi^2\nu^3}~ \Om_G(\nu) ,~~~~~~
\nu=\om/2\pi ,
\label{411}
\eeq
(see Appendix B). As clearly shown in Fig. 4, a sensitivity of
$10^{-25}$ in the kHz range (which is a typical range for present,
Earth-based detectors) would already be enough to enter the
region where we expect a signal, thus constraining (even with a
negative result) the parameters of our class of string cosmology
models\cite{32}. 

\begin{figure}[htb]
   \epsfxsize=6cm
   \centerline{\epsfbox{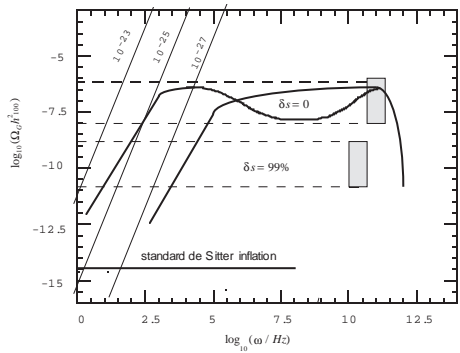}}
   \centerline{\parbox{11.5cm}{\caption{\label{fig:f4}
Position of the peak and of the end point of the relic
graviton spectrum for a typical pre-big bang scenario. }}}
\end{figure}

In fact, any spectrum with a shape that reaches the end point, 
growing not faster than $\om^3$, is in principle allowed in this
context, as the two examples shown by the thick curves in Fig.
4. The slope has to be cubic at low enough frequencies, for
modes crossing the horizon in the dilaton-driven phase. At
higher frequencies the spectrum has to be on the average
non-decreasing. However, it could also be non-monotonic, if 
$\dot\phi$  does not follow exactly, during the string phase, 
 the constant
behaviour of the spacetime curvature, but decreases or
undergoes small oscillations around the string scale (see the
next section). Interestingly enough,, an oscillating
spectrum at high frequency seems to be possible even for a
monotonic evolution of the dilaton background\cite{37}. 

Finally, it may be interesting to point out that for a long enough
string phase it seems possible, in principle, to have a spectrum
that is normalized to the end point value at high frequency, and
which is nevertheless strong enough at the COBE scale to be
significantly constrained by the present data on the CMB
anisotropy. 

Suppose in fact that all scales inside our present horizon
($~\gaq~ \om_0\sim 10^{-18}$ Hz) crossed the horizon during
the high curvature string phase, and let us parametrize the
present (unknown) distribution of tensor perturbations with a
spectral index $n$, such that $n=1$ corresponds to the flat,
scale-invariant spectrum. For all scales reentering the horizon
after the time of matter--radiation equilibrium,
$\om<\om_{eq}\sim 10^{-16}$Hz, there is an additional effect of
parametric amplification due to the matter-dominated phase,
which imposes an extra $\om^{-2}$-dependence on the
spectrum\cite{30b,38}.  By choosing, from eqs. (\ref{47}) and
(\ref{49}), the typical values $\om_1\sim 10^{11}$Hz and
$\Om_G(\om_1)\sim 10^{-6}$ for illustrative purpose, we then 
have the possible normalized distribution
\bea
\Om_G(\om_1, t_0) &=& 10^{-6}\left(\om\over
\om_1\right)^{n-1}, ~~~~~~~~~~~~~~~~~~~~~\om_{eq}<\om<\om_1
\nonumber \\
&=& 10^{-6}\left(\om\over
\om_1\right)^{n-1}\left(\om_{eq}\over \om\right)^2,
~~~~~~~~~\om_{0}<\om<\om_{eq}. 
\label{412}
\eea

For the dilatonic branch of the spectrum the spectral index is
$n=4$, and $\Om_G$ is consequently negligible at the present
horizon scale $\om_0$. For the string branch of the spectrum,
however, $n$ is unknown, and can be bounded from below by
imposing the COBE normalization, which
implies\cite{28} $\Om_G(\om_0, t_0)\sim 10^{-10}$. This
condition is perfectly compatible with the end-point
normalization, and gives $n =37/29\simeq 1.27$. The
corresponding spectral behaviour in the range
$\om_0<\om<\om_{eq}$, namely  $\Om_G\sim \om^{-1.73}$, is not flat 
enough to match the observed CMB anisotropy (a gravity wave 
interpretation of such an isotropy is problematic also in the
standard inflationary context\cite{40}). However, for
$\om>\om_{eq}$, the resulting, slightly growing, ``blue" spectrum
is well consistent with the bound obtained from pulsar timing
data\cite{39},  $\Om_G(\om_p)~\laq ~10^{-8}$ at $\om_p \sim
10^{-8}$Hz, as illustrated in Fig. 5. 

\begin{figure}[htb]
   \epsfxsize=6cm
   \centerline{\epsfbox{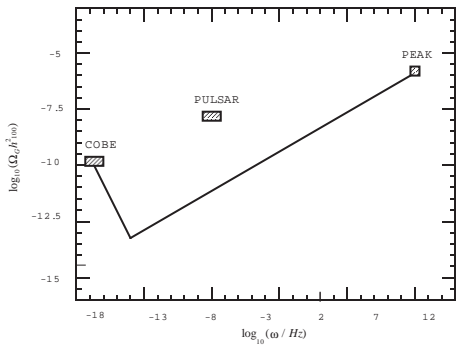}}
   \centerline{\parbox{11.5cm}{\caption{\label{fig:f5}
Example of possible graviton spectrum which has the correct
peak normalization at high frequency, and that satisfies also the
COBE normalization at the frequency scale of our present 
horizon. }}}
\end{figure}

Note that, by assuming the normalization at the COBE scale as
input, since the end-point normalization is fixed, 
we may {\em predict} a spectrum with $\Om_G\sim
10^{-8}$ in the kHz range, which seems to be accessible to the
sensitivity of gravity-wave detectors of the second generation
(see Section 4.5). 

\subsection{A diagrammatic approach to the perturbation
spectrum}
\noindent
At this point, two remarks are in order. The first concerns the
possibility of constraining the graviton spectrum through the
associated amplification of electromagnetic fluctuations, which
in our context is always expected to accompany graviton
production. The second concerns the possible qualitative
modification of the spectrum, in models of background more
complicated than the minimal one previously considered. 

A complete discussion of these points would require a detailed
knowledge of the kinematics and of the higher-order corrections
during the string phase. In the absence of such knowledge, I will
approximate the background with a finite number of phases,
each of them characterized by a different type of power-law
evolution. Also, I will use the lowest-order perturbation
equation, working however only in the asymptotic regimes
$|k\eta|<<1$ and $|k\eta|>>1$ (also called, sometimes
improperly, outside and inside the horizon), assuming that the
qualitative behaviour of the asymptotic solution is not spoiled
by higher-order corrections (see the previous section for
arguments supporting this assumption).

Under the above assumptions, the graviton spectrum can be
conveniently obtained through a sort of ``diagrammatic"
technique, in which one has only to specify the power (in
conformal time) of the scale factor of the various phases, and
the time of horizon crossing of the frequency band considered.

This technique is based on the fact that the exact solution of the
perturbation equation (\ref{42}), for a generic power-law
evolution of the Einstein frame scale factor, $a\sim|\eta|^\a$,
can be written in terms of the first-kind ($H_\nu^{(1)}$) and
second-kind ($H_\nu^{(2)}$) Hankel functions, of index
$\nu=|\a-1/2|$, as follows:
\beq
a=|\eta|^\a \Rightarrow
\psi_k= \left|\eta_k\right|^{1/2}\left[A_+(k)H_\nu^{(2)}(|k\eta|)
+A_-(k)H_\nu^{(1)}(|k\eta|)\right], ~~~~\nu=|\a-1/2|
\label{413}
\eeq
($A_\pm$ are integration constants). When matching the
solution from a phase with power $\a_2$ to a phase with power
$\a_1$, at the time $\eta=\eta_1$,
\beq
\psi_k=\psi_k^2(A^2_\pm, \nu_2), ~~~~\eta<\eta_1 ; ~~~~~~
\psi_k=\psi_k^1(A^1_\pm, \nu_1), ~~~~\eta>\eta_1 ;
\label{414}
\eeq
we may thus easily distinguish the two asymptotic regimes 
$|k\eta_1|<<1$ and $|k\eta_1|>>1$, corresponding respectively
to the comoving frequency of a mode that hits and 
one that does not 
hit the potential barrier of eq. (\ref{42}), whose height
at $\eta_1$ is $|V_1|^{1/2}\sim |\eta_1|^{-1}$. In the first case, 
by using the small argument limit of the Hankel
functions\cite{41}, the continuity of $\psi$ and $\psi'$ gives
\bea
\left(A^1_++A^1_-\right)&=&b_1\left(A^2_++A^2_-\right)
x_1^{\nu_2-\nu_1}+ b_2\left(A^2_+-A^2_-\right)
x_1^{-\nu_2-\nu_1} \nonumber\\
\left(A^1_+-A^1_-\right)&=&b_3\left(A^2_++A^2_-\right)
x_1^{\nu_2+\nu_1}+ b_4\left(A^2_+-A^2_ -\right)
x_1^{-\nu_2+\nu_1},
\label{415}
\eea
where $x_1=|k\eta_1|$ and $b_1,b_2,b_3,b_4$ are complex
numbers with modulus of order $1$ (for $\nu=0$,  $x^\nu$ has to
be replaced by $\ln x$).  In the second case, $|k\eta|>>1$, the
perturbation is approximately unaffected by the background
transition, and the large argument limit of the Hankel functions
gives
\beq
A_+^1=A_+^2, ~~~~~~~~~~~~~ A_-^1=A_-^2 .
\label{416}
\eeq

Suppose now to approximate the background evolution with
$n+1$ different phases, with powers $\a_1, \a_2, ..., \a_{n+1}$,
and solutions $\psi_1,\psi_2, ..., \psi_{n+1}$ of the perturbation
equations, with the corresponding Bessel index $\nu_1,\nu_2, ...,
\nu_{n+1}$. The transitions occur, respectively, at the times
$\eta_n<\eta_{n-1}<...<\eta_2<\eta_1$, and the height of the
effective potential at the transition time $\eta_i$ is
$|V_i|^{1/2}\sim |\eta_i|^{-1}\equiv k_i$ (see Fig. 6).
Normalizing the initial solution to a vacuum fluctuation spectrum,
\beq
A_+^{n+1}=1, ~~~~ A_-^{n+1}=0, ~~~~~~
\psi_{n+1}=|\eta|^{1/2}H^{(2)}_{\nu_{n+1}}(|k\eta|) , 
\label{417}
\eeq
the Bogoliubov coefficient determining the expectation
number of particles, appearing in the final spectral distribution
$\Om_G$ (see eq.(\ref{37a})), is then given by $c_-(k)= A_-^1(k)$. 

\begin{figure}[htb]
   \epsfxsize=6cm
   \centerline{\epsfbox{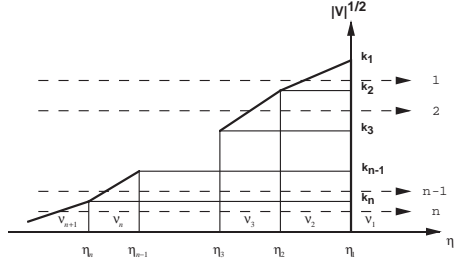}}
   \centerline{\parbox{11.5cm}{\caption{\label{fig:f6}
A background evolution parameterized by $n+1$ phases is in
general associated to a spectrum with $n$ different branches.
The corresponding Bogoliubov coefficients are given in eq.
(4.22). }}}
\end{figure}

In the computation of $A^1_\pm$ we can easily identify $n$
different branches of the spectrum, $(1), (2), ..., (n)$ (see Fig. 6),
corresponding to $n$ asymptotic frequency bands, which only
differ for the power characterizing the scale factor at the time
of horizon crossing. By applying, step by step, the matching
conditions (\ref{415}) and (\ref{416}), one can easily compute the
Bogoliubov coefficients $c_1, c_2, ..., c_n$, for the frequency
bands that are affected, respectively, by $1,2,...,n$ background
transitions. Neglecting numerical coefficients of the order of unity,
we have
\bea
|c_1| &\sim& x_1^{-\nu_1}x_1^{-\nu_2} ,
~~~~~~ k_2<k<k_1
\nonumber \\
|c_2| &\sim& x_1^{-\nu_1}\left(x_1\over
x_2\right)^{-\ep_2\nu_2}x_2^{-\nu_3} ,
~~~~~~ k_3<k<k_2\nonumber \\
|c_3| &\sim& x_1^{-\nu_1}\left(x_1\over
x_2\right)^{-\ep_2\nu_2}\left(x_2\over
x_3\right)^{-\ep_3\nu_3}x_3^{-\nu_4} ,
~~~~~~ k_4<k<k_3\nonumber \\
......&\sim&..........................
........................................... \nonumber\\
|c_n| &\sim& x_1^{-\nu_1}\left(x_1\over
x_2\right)^{-\ep_2\nu_2}\left(x_2\over
x_3\right)^{-\ep_3\nu_3}......\left(x_{n-1}\over
x_x\right)^{-\ep_n\nu_n}x_n^{-\nu_{n+1}}, ~k<k_n
\label{418}
\eea
where
\beq
x_i=|k\eta_i|, ~~~~~~~~ \ep_i=
{\rm sign}\ln \left(x_i\over x_{i-1}\right), ~~~~~~~~
i=1,2,...,n.  
\label{419}
\eeq
The $\ep_i$ coefficients have been inserted to take into account
the possibility 
of a non-monotonic potential (unlike that of Fig. 6), since for
$x_i<x_{i-1}$ the ratio $x_{i-1}/x_i$ enters  the Bogoliubov
coefficients with the opposite power. Note also that a phase
with vanishing effective potential, like radiation-dominated
evolution, $\a_i=0$, has a Bessel index $\nu_i=1/2$. 

As a simple application of the general prescription (\ref{418}),
let us write down the Bogoliubov coefficients for a background
characterized by four different phases and by a non-monotonic
effective potential, which has a first break at $\eta_s$, a peak
at $\eta_1$, and which becomes radiation-dominated at $\eta_r$,
with 
$x_s>x_r>x_1$ (see Fig. 7). The Bessel indices are,
respectively, $\nu_0, \nu_s,\nu_1 $ and $\nu_r=1/2$. 
The graviton spectrum is characterized by three branches,
differing by the number of background transitions by which a
given mode is affected. The corresponding Bogoliubov
coefficients can be immediately deduced from the diagram of
Fig. 7, by applying the general prescription (\ref{418}):
\bea
|c_1| &\sim& x_1^{-\nu_1}x_1^{-\nu_s} ,
~~~~~~~~~~~~~~~~~~ k_r<k<k_1
\nonumber \\
|c_2| &\sim& x_r^{-1/2}\left(x_r\over
x_1\right)^{\nu_1}x_1^{-\nu_s} ,
~~~~~~~~~ k_s<k<k_r\nonumber \\
|c_3| &\sim& x_r^{-1/2}\left(x_r\over
x_1\right)^{\nu_1}\left(x_1\over
x_s\right)^{-\nu_s}x_s^{-\nu_0} ,
~~~~k<k_s
\label{420}
\eea
A similar type of background will be discussed in Section 4.4. 

\begin{figure}[htb]
   \epsfxsize=6cm
   \centerline{\epsfbox{f7.epsf}}
   \centerline{\parbox{11.5cm}{\caption{\label{fig:f7}
Example of background with non-monotonic potential.
. }}}
\end{figure}

As discussed in the previous section,  the square of the
Bogoliubov coefficient determines the expectation number of
the produced gravitons and the spectral distribution $\Om_G$, 
according to eqs. (\ref{37}) and (\ref{37a}). In critical units,
\beq
\Om_G(\om,t)={\om\over \r_c}{d\r\over d\om}=
{8\om^4 \over 3\pi M_p^2H^2(t)}|c_-|^2 .
\label{421}
\eeq
By expressing all ratios in terms of proper frequencies, 
$x_i/x_j\equiv |k\eta_i|/|k\eta_j|\equiv k_j/k_i=\om_j/
\om_i$, and referring the spectrum to the proper frequency
 of the final transition, $\om_1(t)=k_1/a(t)\sim H_1a_1/a$, we
easily obtain, from eqs. (\ref{418}) and (\ref{421}), the spectral
distribution $\Om(\om)$ corresponding to the $n$ branches of
Fig. 6:
\bea
\Om_1 &\sim& \left(H_1\over M_p\right)^2
\left(H_1\over H\right)^2\left(a_1\over a\right)^4
\left(\om\over \om_1\right)^{4-2\nu_1-2\nu_2},
~~~\om_2<\om<\om_1 \nonumber\\
\Om_2 &\sim& \left(H_1\over M_p\right)^2
\left(H_1\over H\right)^2\left(a_1\over a\right)^4
\left(\om\over \om_1\right)^{4-2\nu_1-2\nu_3}
\left(\om_2\over \om_1\right)^{-2\ep_2\nu_2+2\nu_3},
~~~\om_3<\om<\om_2 \nonumber\\
\Om_3 &\sim& \left(H_1\over M_p\right)^2
\left(H_1\over H\right)^2\left(a_1\over a\right)^4
\left(\om\over \om_1\right)^{4-2\nu_1-2\nu_4}
\left(\om_2\over \om_1\right)^{-2\ep_2\nu_2}
\left(\om_3\over \om_2\right)^{-2\ep_3\nu_3}
\left(\om_1\over \om_3\right)^{-2\nu_4},\nonumber\\
~&~&~\om_4<\om<\om_3 \nonumber\\
...&\sim&...................................
.................................................\nonumber \\
\Om_n &\sim& \left(H_1\over M_p\right)^2
\left(H_1\over H\right)^2\left(a_1\over a\right)^4
\left(\om\over \om_1\right)^{4-2\nu_1-2\nu_{n+1}}
\left(\om_2\over \om_1\right)^{-2\ep_2\nu_2}
\left(\om_3\over \om_2\right)^{-2\ep_3\nu_3}
....\nonumber\\
....&~&\left(\om_n\over \om_{n-1}\right)^{-2\ep_n\nu_n}
\left(\om_1\over \om_n\right)^{-2\nu_{n+1}}, 
~~~~~~~~~~~~~~~~ \om<\om_n .
\label{422}
\eea

It is important to point out that the slope of the spectrum, for a
given branch $\Om_i$, is sensitive to the particular time
evolution of the background only when hitting and when leaving
the barrier: in fact, $\Om_i (\om)\sim
\om^{4-2\nu_1-2\nu_{i+1}}$. Note also that, if we wish to refer 
the spectrum to a frequency $\om_k$ other than $\om_1$ (for
instance to the peak frequency, which may be different from $\om_1$
if the potential is non-monotonic), we have to simply multiply
the $i$-th branch of the spectrum by
$(\om_k/\om_1)^{4-2\nu_1-2\nu_{i+1}}$.

Let us now apply the general prescription (\ref{422}) to a typical
pre-big bang scenario, in which the background evolves from an
initial dilaton-driven phase with $\nu_{n+1}=0$. Eventually, the
background becomes radiation-dominated, so that $\nu_1=1/2$,
and the ratio
\beq
\left(H_1\over H\right)^2\left(a_1\over a\right)^4=
\Om_r(t)
\label{423}
\eeq
gives the radiation energy density, in critical units, rescaled to a
time $t>t_1$. The various phases from $\eta_n$ to $\eta_1$
correspond to possible types of background during the string
era, and the ratio
\beq
g_1=H_1/M_p
\label{424}
\eeq
is taken to be 
of the same order as the present value of the fundamental
ratio  between string and Planck mass, $g_1\sim M_s/M_p\sim
0.3$ -- $0.03$. Let us assume, also, that the 
evolution of the metric background, in the Einstein frame, 
is always accelerated from
$\eta_n$ to $\eta_1$, so that the potential $|V(\eta)|$ is
non-decreasing, $\om_1$ is also the maximal amplified frequency, and
$\ep_i=1$ for all the transitions. 
The spectral distribution (\ref{422}) for the $i$-th band of
frequency, crossing the horizon during the string phase ($i<n$),
and affected by $i$ background transitions, can thus be
written
\bea
\Om_i(\om)&\sim&
g_1^2~\Om_r(t)\left(\om\over\omega_1\right)^{3-2\nu_{i+1}}
\left(\om_2\over\omega_1\right)^{-2\nu_{2}}
\left(\om_3\over\omega_2\right)^{-2\nu_{3}}.... \nonumber\\
&~&.....
\left(\om_i\over\omega_{i-1}\right)^{-2\nu_{i}}
\left(\om_1\over\omega_i\right)^{-2\nu_{i+1}}, ~~~~~~~~~
\om_{i+1}<\om<\om_i .
\label{425}
\eea

As anticipated in the previous section, since $\nu_{i+1}\geq 0$, this
spectrum is characterized by a maximal slope $\Om\sim \om^3$
(which is the same slope as that of the frequency band crossing the
horizon in the dilaton-driven phase, modulo log corrections). 
Also, since
$\om_n<\om_{n-1}<...<\om_2<\om_1$, the peak of the spectrum, 
$g_1^2\Om_r$, is reached at $\om=\om_1$, so that the spectral 
distribution 
has to be on the average non-decreasing from $\om_n$ to
$\om_1$, i.e. $\Om_i\leq \Om_1$. However, the spectrum could
be non-monotonic during the string phase (in spite of the
monotonicity of the potential). Consider in fact the ratio
\beq
{\Om_i(\om_i)\over \Om_{i-1}(\om_{i-1})}=
\left(\om_i\over \om_{i-1}\right)^{3-2\nu_i} . 
\label{426}
\eeq
As $\om_i<\om_{i-1}$, the spectrum may be decreasing in the
frequency band $\om_i<\om<\om_{i-1}$, provided
$3<2\nu_i=|2\a_i-1|$.  If we assume a constant curvature
evolution in the String frame, the decreasing  slope 
may be due to a phase of decreasing dilaton.

Suppose in fact that, during the string phase, there is a period 
in which the scale factor and the dilaton evolve in time 
(in the String frame) as
\beq
\ti a_i=(-\eta)^{-1}, ~~~~~~~~ \ti \phi_i =\ga \ln \ti a_i,
~~~~~~~~\ga <0
\label{430a}
\eeq
(corresponding to $\ti H_i=$const and decreasing dilaton, 
${\dot{\ti \phi_i}}=\ga \ti H_i <0$). Tranforming to the Einstein
frame\cite{2,3} we have
\beq
a_i=\ti a_ie^{-\phi_i/2}=\left(-\eta\right)^{-1+\ga /2} , ~~~~~~~~
\phi_i=\ti \phi_i= -\ga\ln (-\eta)
\label{430b}
\eeq
(conformal time is the same in the two frames). The slope of the
corresponding frequency band is then decreasing, since 
\beq
3-2\nu_i=3-|2\a_i-1| =3+2\left({\ga \over 2}-1\right)-1= 
\ga <0 .
\eeq

\subsection{Constraints from photon production}
\noi
In the context of the pre-big bang scenario, the time variation of
the background fields that amplifies the perturbations of the
metric tensor also amplifies the quantum fluctuations of the
electromagnetic field\cite{24,41a}. The electromagnetic
perturbation equation has the same form as eq. ({\ref{42}), with
the only difference that the effective potential is generated by
a time-dependent dilaton field, 
\beq
V(\eta)= {\phi'^2\over 4} -{\phi''\over 2}.
\label{427}
\eeq
For a cosmological evolution that can be approximated by
various phases, characterized by a logarithmic variation of the
dilaton, $\phi_i=-\b_i\ln|\eta|$, the electromagnetic
perturbation spectrum can thus be computed with the
diagrammatic technique illustrated in the previous section. The
result is a spectrum with the same general structure as that
of the graviton spectrum, but with different (in general) values
of the Bessel indices $\nu_i$. The phenomenological bounds
imposed on electromagnetic perturbations constrain the
parameters of the cosmological background and then, indirectly,
also the unknown slope of the graviton spectrum during the
string phase. 

In order to discuss this effect we shall work in the context of
the minimal model of Section 4.1, characterized by three phases
(dilaton $\ra$ string $\ra$ radiation), and we shall parametrize
the unknown background evolution during the string phase (in
the Einstein frame) with two powers $\a$ and $\b$,
\beq
a\sim |\eta|^\a, ~~~~~~~\phi\sim -2\b \ln|\eta|, ~~~~~~~
\eta_s<\eta<\eta_r .
\label{428}
\eeq
For the dilaton-driven phase ($\eta<\eta_s$) and for the
radiation-dominated phase ($\eta>\eta_r$) the background
evolution is known\cite{2,3}, and is given respectively by
\bea
a\sim |\eta|^{1/2}&,& ~~~~~~~\phi\sim -\sqrt 3 \ln|\eta|,
~~~~~~~ \eta<\eta_s , \nonumber\\
a\sim |\eta|&,& ~~~~~~~\phi\sim {\rm const},
~~~~~~~~~~~~~~ \eta>\eta_r .
\label{429}
\eea
The effective potential $|V(\eta)|$ grows monotonically for
$\eta<\eta_r$ and goes rapidly to zero for $\eta>\eta_r$, both
for gravitons and photons. The spectra have two branches, which 
can  easily be deduced from the diagram of Fig. 8. 

\begin{figure}[htb]
   \epsfxsize=6cm
   \centerline{\epsfbox{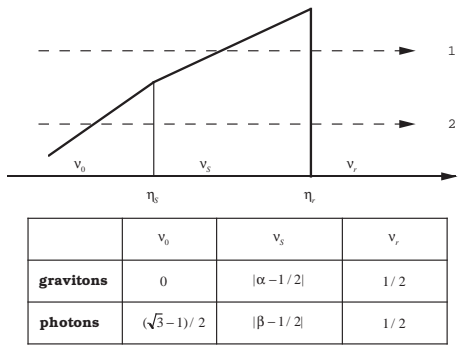}}
   \centerline{\parbox{11.5cm}{\caption{\label{fig:f8}
Diagram and Bessel indices for the graviton and photon spectrum
in a minimal, two-parameter model of background. }}}
\end{figure}

The Bessel indices $\nu_i$ are fixed by the time evolution of the
background, according to eqs. (\ref{428}) and (\ref{429}). The
two parameters $\a$ and $\b$ are not independent, however, if
we require\cite{24,27} (as  seems natural) that, during the
string phase, the curvature remains controlled by the string scale
$M_s=\la_s^{-1}$. This implies, in the Einstein frame\cite{35a},
\beq
\left|H_s\over H_1\right|\simeq
\left|\eta_r\over \eta_s\right|^{1+\a}\simeq
{g_s\over g_r}\simeq
\left|\eta_r\over \eta_s\right|^{\b}, ~~ g_s=e^{\phi_s/2}, ~
g_r=e^{\phi_r/2}, 
\label{431}
\eeq
from which
\beq
1+\a \simeq \b \simeq -{\log (g_s/g_r)\over \log
|\eta_s/\eta_r|} . 
\label{431}
\eeq
Defining the convenient parameters
\bea
x &=& \log_{10}
|\eta_s/\eta_r|=\log_{10} z_s , \nonumber\\
y&=& \log_{10}
(g_s/g_r)=\log_{10} {H_s/M_p\over H_r/M_p} ,
\label{432}
\eea
related respectively to the duration of the string phase, $z_s=
|\eta_s/\eta_r|$, and to the corresponding shift of the dilaton
coupling, $g_s/g_r$, we obtain
\beq
\a-{1\over 2}={3\over 2}+{y\over x}, ~~~
\b-{1\over 2}={1\over 2}+{y\over x}, ~~~
{\om_r\over \om_s}={H_r a_r\over H_s
a_s}\simeq\left|\eta_s\over \eta_r\right|=10^x. 
\label{433}
\eeq
By applying the rules of the previous section we can now
parametrize the spectra in terms of $x$ and
$y$. For the two branches of the graviton spectrum we obtain,
from Fig. 8,
\bea
\Om_G &\sim &
g_1^2~\Om_r(t)\left(\om\over\omega_1\right)^{3-|3+2y/x|},
~~~~~10^{-x}<{\om\over \om_1}<1 \nonumber\\
&\sim &g_1^2~\Om_r(t)\left(\om\over\omega_1\right)^{3}
10^{|3x+2y|},
~~~~~{\om\over \om_1}<10^{-x}
\label{434}
\eea
(we have chosen to refer all frequencies to the maximal
amplified one, $\om_1$). With the same rules we obtain the two
branches of the electromagnetic spectrum,
\bea
\Om_{em} &\sim &
g_1^2~\Om_r(t)\left(\om\over\omega_1\right)^{3-|1+2y/x|},
~~~~~~~~~~10^{-x}<{\om\over \om_1}<1 \nonumber\\
&\sim &g_1^2~\Om_r(t)\left(\om\over\omega_1\right)^{4-\sqrt 3}
10^{x(1-\sqrt 3)+|x+2y|},
~~~~~{\om\over \om_1}<10^{-x}.
\label{435}
\eea

The back reaction of the electromagnetic perturbations on the
geometry turns out to be 
negligible, consistently with the use of a
homogeneous and isotropic background, provided we
impose on the two-dimensional parameter space of our
model the condition\cite{24} $y\gaq -2x$. Also, the amplified
perturbations are large enough to act as ``seeds" for the
galactic magnetic field, provided\cite{42} $\Om_G(10^{-14}
{\rm Hz})~\gaq~ 10^{-34}\Om_r$. These two conditions define an
allowed region in the $\{x,y\}$ plane, and constrain the possible
slope of the string branch of the graviton spectrum.

The first condition, by itself, does not provide significant
constraints. When combined with the second, however, we
obtain the constraint
\beq
2<3-\left|3+2{y\over x}\right|<4, 
\label{436}
\eeq
which for the graviton spectrum (\ref{434}) defines the allowed
region shown in Fig. 9 (where I have used $g_1=0.1$ and
$\om_1=10^{11}$Hz). The bold dashed line in Fig. 9 represents the
prediction for the graviton spectrum for the case that the
amplified electromagnetic perturbations are strong enough, on a
large scale, to contribute in a significant way to the observed
CMB anisotropy\cite{26}. This possibility would require indeed a
flat enough string branch of the electromagnetic spectrum,
$\Om_{em}\sim \om^\ep$, $0<\ep<1$, and could be implemented
consistently with other phenomenological constraints only for a
background in which the comoving amplitude of perturbations
grows in time also outside the horizon\cite{26}. 
In terms
of our parameters this gives the conditions 
$y>-x/2$ and $y/x=1+\ep$ and implies, for 
the string branch of the spectrum, a 
slope $\Om_G\sim \om^{2+\ep}$.  

\begin{figure}[htb]
   \epsfxsize=7cm
   \centerline{\epsfbox{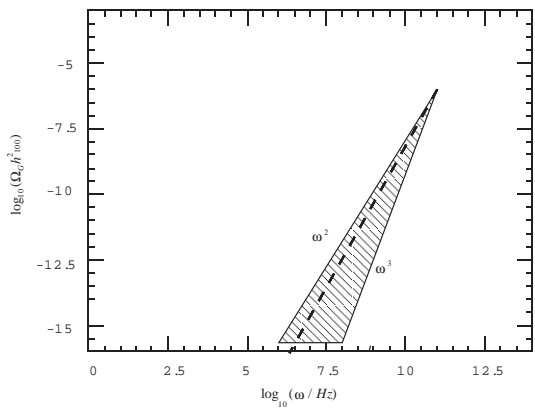}}
   \centerline{\parbox{11.5cm}{\caption{\label{fig:9}
The shaded area shows the allowed region for the graviton
spectrum, in a model in which the electromagnetic fluctuations
are amplified enough to act as ``seeds" for the galactic
magnetic field.   }}}
\end{figure}

It is important to note that the allowed region of Fig. 9
corresponds to a very small energy density of the relic graviton
background in the kHz range, $\Om_G~\laq~10^{-21}$; this is 
certainly outside the sensitivity of 
gravity wave detectors now existing or planned (see Section
4.5). A possible future detection of a relic graviton background in
the kHz range could thus be used to exclude, in our cosmological
scenario, a
 primordial origin of the seed fields and a possible
electromagnetic origin of the CMB anisotropy (at least in the
context of a minimal model, and in the range of validity of the
low-energy perturbation equation). Conversely, if we believe
that the ``dilatonic" amplification of the electromagnetic
fluctuations is efficient enough to be responsible for the above
effects, we should expect a graviton background outside the
detector sensitivities, in the kHz range. This may provide strong
motivations for developing detectors with improved sensitivities
in the very high frequency range ($\sim$ GHz), near the end point
of the spectrum, where the background is expected to reach the
peak intensity. At lower frequencies, the absence of a signal at
the sensitivity 
level of $10^{-6}$  (or better) in $\Om_G$ would 
nevertheless be important, since it would support, indirectly, a
primordial dilatonic origin of the cosmic magnetic fields. 

\subsection{Non-minimal models}
\noi
In the minimal model of pre-big bang assumed in the previous
Sections, the end of the high curvature string phase was nearly
coincident with the beginning of the radiation-dominated,
constant dilaton phase. It is not impossible, however, to imagine
more complex models of backgrounds, in which the dilaton is
still growing while the curvature starts decreasing and where the
radiation era with frozen dilaton starts at much later times,
when the curvature scale is much below the string scale. An
example of such a background is illustrated in Fig. 10 (the
minimal model corresponds to the particular case $\eta_r\sim
\eta_1$, and $g_1\sim 1$). 

\begin{figure}[htb]
   \epsfxsize=6cm
   \centerline{\epsfbox{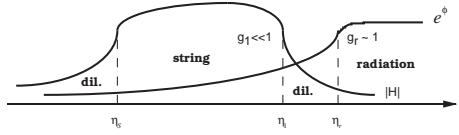}}
   \centerline{\parbox{11.5cm}{\caption{\label{fig:f10}
Non-minimal model of pre-big bang background. }}}
\end{figure}

It is conceivable, in fact, that the combined action of loops and
$\ap$ corrections may stop the growth of the curvature and
may induce a  change of branch of the solution, while the string
coupling remains growing\cite{7b}. Only later could the dilaton 
be attracted to a fixed point of the non-perturbative potential.
In that context, however, the amplified vacuum fluctuations are
not expected to become naturally dominant, and the radiation
should arise from other sources (for instance, coherent dilaton
oscillations). 

If we accept a model like that of Fig. 10, the non-vanishing part
of the effective potential appearing in the perturbation equation
(\ref{42}) is no longer monotonic. Consequently, high-frequency 
modes may reenter the horizon (or leave the potential barrier)
{\em before} the beginning of the radiation era, i.e. for
$\eta_1<\eta<\eta_r$. This modifies the slope of the
spectrum in the high frequency sector. For the background of
Fig. 10, in particular, the spectrum has three branches, whose
slope depends on the relative duration of the string phase and of
the new phase inserted before the radiation era. If
$|\eta_r/\eta_1|<|\eta_s/\eta_1|$, i.e. $\om_r>\om_s$, the
spectrum corresponds to the diagram of Fig. 7. If
$\om_r<\om_s$ the three branches of the spectrum are instead
obtained from the diagram of Fig. 11, where I have supposed
that the decelerated phase from $\eta_1$ to $\eta_r>0$ is again
a dilaton-driven solution of the lowest order effective action,
such that
\beq
a\sim \eta^{1/2}, ~~~~ e^\phi \sim \eta^{\sqrt 3}, ~~~~
\left(H_1\over H_r\right)\simeq \left(g_r\over g_1\right)^{\sqrt
3}\simeq \left(a_r\over a_1\right)^3, ~~~~
\eta >0 .
\label{437}
\eeq
By referring the spectrum to the maximal amplified frequency
$\om_1$, and applying the rules of Sect. 4.2, we obtain from 
Fig. 11
\bea
\Om_1 &\sim &
g_1^2\left(g_r\over g_1\right)^{2/\sqrt 3}
\Om_r(t)\left(\om\over\omega_1\right)^{4-2\nu_s},
~~~~~~~~~\om_s<\om<\om_1 \nonumber\\
 &\sim &
g_1^2\left(g_r\over g_1\right)^{2/\sqrt 3}
\Om_r(t)\left(\om\over\omega_1\right)^{4}
\left(\om_s\over \om_1\right)^{-2\nu_s},
~~~~~~~~\om_r<\om<\om_s \nonumber\\
 &\sim &
g_1^2\left(g_r\over g_1\right)^{2/\sqrt 3}
\Om_r(t)\left(\om\over\omega_1\right)^{3}
\left(\om_s\over \om_1\right)^{-2\nu_s}
\left(\om_r\over \om_1\right),
~~~~\om<\om_s
\label{438}
\eea
(neglecting logarithmic corrections, and using the relation
$\om_1/\om_r=(H_1/H_r)^{2/3}$, valid for the background
(\ref{437})).

\begin{figure}[htb]
   \epsfxsize=6cm
   \centerline{\epsfbox{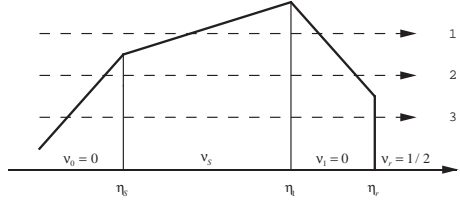}}
   \centerline{\parbox{11.5cm}{\caption{\label{fig:f11}
The three branches of the graviton spectrum for the background of 
Fig. 10, and for the case $\om_r<\om_s$. }}} 
\end{figure}

For the case $\om_s<\om_r$ the spectrum is similarly obtained
from the diagram of Fig. 7. This spectrum differs from that of
the minimal model in three important aspects:

\begin{itemize}
\item{}The end point value of the spectrum, $\Om_1(\om_1)$, is
shifted to arbitrarily low values $\sim \Om_r g_1^{2(1-1/\sqrt
3)}<<\Om_r$, since $g_1<<g_r\sim 1$. The maximal amplified
frequency is correspondingly shifted,
\beq
\om_1(t_0)={\om_1\over\om_r}{H_ra_r\over a(t_0)}\sim g_1^{1/2}
T_0 \left(H_1\over H_r\right)^{1/6}\sim g_1^{1/2}\left(g_r\over
g_1\right)^{1/2\sqrt 3}\times 10^{11}{\rm Hz} <<
 10^{11}{\rm Hz}
\label{439}
\eeq
($T_0$ is the present CMB temperature).

\item{}
The maximal allowed slope is steeper, $\Om\sim \om^4$ instead
of $\Om \sim \om^3$. 

\item{}The spectrum may be truly non-monotonic, with a peak
value at $\om_s$ higher than the end point value,
$\Om(\om_s)>\Om(\om_1)$, provided $4<2\nu_s$.
 \end{itemize}

So, in spite of the fact that the end point is shifted to lower
values, the relic background corresponding to a non-minimal
model could be detected even more easily than in the minimal
case, because of a possible peak at lower frequencies, more accessible
to present detectors. It is true that the upper limit, 
$\Om_g\sim
10^{-2}\Om_r$, cannot be obtained naturally in a non-minimal
scenario. However, the analysis of photon production in this
scenario shows that such a maximal peak signal would not be
excluded by the bounds on the electromagnetic spectrum, provided
$\om_r<\om_s$, i.e. provided the post-big bang dilaton-driven
phase is long enough with respect to the string phase. It is thus
important to point out that, in the context of a non-minimal
model, the possibility is not excluded to see a graviton background  
with a peak $\Om_G\sim10^{-6}$ at an arbitrarily low frequency
$\om_s$ ($>10^{-8}$Hz, however, because of the pulsar
bound\cite{39}), without implying that the spectrum is flat from
$\om_1$ to $\om_s$, thus evading 
the otherwise constraining bound on the total integrated
energy following from nucleosynthesis (see eq. (\ref{410})). 

\subsection{Experimental sensitivities}
\noi
It may be interesting, at this point, to compare the relic graviton
background expected in the context of the pre-big bang scenario
with the experimental sensitivity of gravity wave detectors. A
precise and complete discussion of the present experimental
possibilities of detecting a stochastic gravity wave background
is, of course, outside the scope of this paper (see for instance
the reviews of Refs. [52,53]). Here I will simply report, at the
level of an order of magnitude estimate, the sensitivities accessible
to some detectors that are already existing and operating, and
some other that are now under construction
and  will be operating in the near future.

The ``theory versus experiments" situation is 
qualitatively sketched in Fig.
12, where the bold upper border line defines the maximal
allowed region for the relic graviton spectrum in a string
cosmology context, obtained from Fig. 4; the bold lower line
gives the corresponding border in the standard inflationary
context. 

\begin{figure}[htb]
   \epsfxsize=6cm
   \centerline{\epsfbox{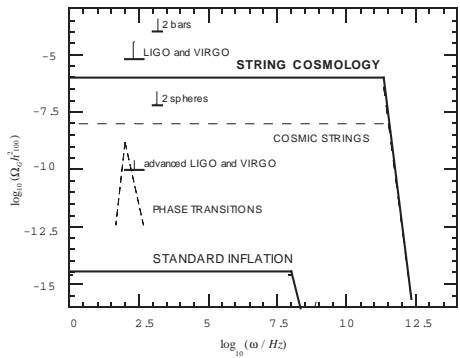}}
   \centerline{\parbox{11.5cm}{\caption{\label{fig:f12}
Present and future experimental sensitivities, compared with
the allowed region for the graviton spectrum. }}} 
\end{figure}

Available data, at present, come from the cryogenic, 
resonant-mass detector EXPLORER, operating at CERN, and
provide the upper bound\cite{45} $\Om_G~\laq~ 500$ at a
frequency $\nu=923~$Hz, which is too high to be significant for a
relic cosmological spectrum. However, by cross-correlating the
data obtained from two bar detectors with the same
characteristics as those already existing (EXPLORER, NAUTILUS,
AURIGA), it will be possible to reach, in very few years, the
level of sensitivity\cite{45} $\Om_G\sim 8\times 10^{-4}$ around
the kHz range. This is still outside, but not so far off, the border
of the interesting region, as shown in Fig. 12. Even better
sensitivities, in the same frequency range, can be reached
through the cross-correlation of a bar and an interferometric
detector\cite{46,47}. And in fact the first generation of
interferometric detectors, LIGO\cite{48} and VIRGO\cite{49},
when operating, will soon reach a sensitivity\cite{43} 
$\Om_G\sim  10^{-5}$ centred around the $100~$Hz frequency
band. The possible improvement in the sensitivity of 
 microwave cavities available at present, operating as gravity wave
detectors, is also under active study\cite{50}.

Also shown in Fig. 12 are the sensitivities expected from the
cross-correlation of two resonant spherical
detectors\cite{45,47}, $\Om_G\sim 10^{-7}$ at $\nu\sim 1~$kHz,
and the planned sensitivity of the advanced interferometric
detectors\cite{43,51}, $\Om_G\sim  10^{-10}$ at $\nu\sim
10^2$kHz. Both are well inside the allowed region, and are thus 
able to constrain significantly, in case of no detection, the
parameters of our class of string cosmology models. To the best
of my knowledge, in this range of frequency there are only two
other possible primordial backgrounds, competing in intensity
with the relics of a pre-big bang phase. The first is the 
background (with very flat spectrum) generated by cosmic
strings\cite{52} or by other topological defects\cite{53},
characterized by $\Om_G\sim 10^{-8}$. The second is a
background sharply peaked around $10^2-10^3$Hz, generated by first
order phase transitions\cite{52,54} in the context of extended
inflation models, with peak value $\Om_G\sim 10^{-8}$--$10^{-9}$.
Both these two possibilities are illustrated by the dashed curves
of Fig. 12. 

Finally, it should be recalled that for non-minimal models of the 
pre-big bang era, the allowed string cosmology region of Fig. 12 can
be extended to the left, in principle  
without further suppression down to frequencies 
of $10^{-7}$Hz. The space interferometer
LISA\cite{55}, operating around the mHz range, will thus provide
significant bounds on non-minimal models and, at the maximal
expected sensitivity level of $\Om_G\sim 10^{-10}$, also on
minimal models whose spectrum decreases enough to evade
the nucleosynthesis bound. 

\vskip 1 cm

\renewcommand{\theequation}{5.\arabic{equation}}
\setcounter{equation}{0}
\section{Conclusion}
\label{sec:5}
\noindent
The cosmological scenario in which I have discussed the
production of a relic graviton background was based on two main
assumptions: 
\begin{itemize}
\item{}the initial state of the Universe is the string perturbative
vacuum;
\item{}the pre-big bang phase evolves smoothly into a final 
decelerated state, dominated by radiation and with constant
dilaton.
\end{itemize}

While the first assumption represents an arbitrary choice of the
initial conditions, the second one reflects our present ignorance
about string cosmology in the high-curvature and
non-perturbative regime, where the transition from the pre- to
the post-big bang epoch is expected to occur. Our present lack
of knowledge about the kinematics of such a transition forbids,
in particular, a precise prediction of the detailed shape of the
graviton spectrum. 

If we accept the  above two assumptions, however, we can
already conclude that from a phenomenological point of
view the pre-big bang scenario is significantly different from
the standard inflationary scenario, and that these differences
lead to effects that can be tested (at least in principle) even
more easily than the corresponding effects of the standard
scenario. 

In particular, there is no compelling reason in this context to
exclude a relic graviton background as high as 
\beq
\Om_g(\om)\sim 10^{-6}h_{100}^{-2},
\label{51}
\eeq
in the whole range from $1~$Hz to $100$ GHz. Thus, experimental
measurements performed at this level of sensitivity  would be 
true tests of Planck-scale physics. The detection of such a
cosmic graviton background would represent an event of an importance
comparable to the detection of the cosmic black-body 
radiation\cite{56}. Even more important than this, 
in some sense, 
because the CMB photons are relic radiation from the hot, ``big bang"
phase that occurred in the past history of our Universe. Those
gravitons would, instead, be relics of a much earlier ``pre-big
bang" epoch.

\vspace{1cm}
{\it Acknowledgements:\/}
This work is supported in part by EC contract
ERBCHRX-CT94-0488. I am grateful to Ramy Brustein, Massimo
Giovannini and Gabriele Veneziano for useful discussions and for
fruitful collaboration on the relic graviton background in string
cosmology. 

\vskip 1 cm

\renewcommand{\theequation}{A.\arabic{equation}}
\setcounter{equation}{0}
\section{Appendix A \\ The dimensionless amplitude $|\da_h|$}
\label{sec:A}
\noindent
Consider tensor perturbations, in a conformally flat metric
background, and in the transverse traceless gauge,
\beq
g_{\mu\nu}+\da g_{\mu\nu}, ~~~~~~~~
\da g_{\mu\nu}= a^2(\eta) h_{\mu\nu}, ~~~~~~~
h_\mu^\mu=0=\nabla_\nu h_\mu^\nu.
\eeq
The gravitational action, perturbed up to second order in each
of the two physical polarization modes $h(x)$, 
\beq
{M_p^2\over 16\pi}\int d^3x d\eta~ a^2\left[h'^2-(\nabla_i
h)^2\right] ,
\eeq
defines the canonical variable\cite{20a,21,22,57} $u$, which
diagonalizes the perturbed action:
\beq
u=a h M_p ,
\eeq
and whose Fourier components $u_k$ are required to satisfy
canonical commutation relations: 
\beq
\left[u_k,\dot u_{k'}\right] =i\da_{kk'}.
\eeq
This fixes the dimensionality of the mode $h_k$,
$[h_k]=[u_k/M_p]=[k^{-3/2}]$, and gives the correct
normalization for the Fourier transform in a finite volume $V$,
\beq
h(\vec x,t)={1\over \sqrt V}\sum_{\vec k}h_{\vec k}(t) e^{i\vec k
\cdot \vec x} .
\label {a5}
\eeq
In the continuum limit
\beq
h(\vec x,t)={\sqrt V\over (2\pi)^3}\int d^3k~ h(\vec k,t)e^{i \vec k
\cdot \vec x} .
\label {a6}
\eeq

Consider now the two-point correlation function 
\beq
\xi(\vec r)=\langle h(\vec x) h(\vec x+\vec r)\rangle,
\label{a7}
\eeq
where the brackets denote the quantum expectation value on a
given state, if we work in the second-quantization formalism in
which the perturbation field is expanded into annihilation and
creation operators. We  recall, however, that the
perturbation background is a stochastic background as a
consequence of its quantum origin, since it is obtained by
amplifying the quantum fluctuations of the vacuum\cite{58}, for
which 
\beq
\langle a^\dagger_k a_{k'}\rangle= |f|^2\da^3(k-k'),
\label{a8}
\eeq
where $f$ is an appropriate normalization coefficient. This
means that,  if we perform the classical limit by replacing 
quantum expectation values with ensemble averages, we 
are led to the stochastic condition
\beq
\langle h(\vec k) h(-\vec k')\rangle=|h(\vec k)|^2{(2\pi)^3\over
V} \da^3(k-k'). 
\label{a9}
\eeq
For a stochastic background the correlation function (\ref{a7})
thus becomes
\beq
\xi(\vec r)=\int{d^3k\over (2\pi)^3}|h(k)|^2 e^{-i \vec k
\cdot \vec r} = {1\over 2\pi^2}\int{dk\over k} {\sin kr\over kr}
\left|\da_h(k)\right|^2 ,
\label{a10}
\eeq
where we have defined
\beq
\left|\da_h(k)\right|=k^{3/2}|h(k)|. 
\label{a11}
\eeq
Note that we have used the reality condition, $h^\ast(k)=h(-k)$, and
the so-called isotropy condition, i.e. the assumption that $|h(k)|$
is a function only of $k=|\vec k|$. The brackets of
eq. (\ref{a7}) can also be formally defined as a macroscopic
average over space:
\beq
\xi(\vec r)={1\over V}\int d^3x ~h(\vec x) h(\vec x+\vec r);
\label{a12}
\eeq
the result for $\xi(\vec r)$ is exactly the same as in eq.
(\ref{a10}). By evaluating the correlation function for a distance
$r=k^{-1}$ we  finally obtain from (\ref{a10})
\beq
\left[\xi^{1/2}(r)\right]_{r=k^{-1}} \sim \left|\da_h(k)\right|. 
\label{a13}
\eeq
The variable $\left|\da_h(k)\right|$ can thus be interpreted as the 
typical, dimensionless amplitude of tensor perturbations over a
comoving length scale $r=k^{-1}$.

The spectral density of perturbations, $\Om_G(\om)$, can be
conveniently expressed in terms of $ \left|\da_h(k)\right|$ by
recalling that the average energy density, summing over
polarization, is defined by\cite{27a}
\beq
\r_g={dE_g\over a^3 d^3x}={M_p^2\over 8\pi}
\langle \dot h^2\rangle=
-{M_p^2\over a^3 8\pi}{V\over (2\pi)^6}\int d^3kd^3k'\om \om' 
 e^{i\vec x \cdot (\vec k+\vec k')} \langle h(k)h(k')\rangle,
\label{a15}
\eeq
where $ \om= |\vec k|/a$. By using the stochastic condition
(\ref{a9}), in the hypothesis $\left|\da_h(k)\right|=
\left|\da_h(-k)\right|$, 
\beq
{dE_g\over a^3 d^3x d\ln k}={M_p^2\over (2\pi a)^3}
\left|\da_h(k)\right|^2 \om^2. 
\label{a15}
\eeq
The spectral distribution of the proper energy density, in critical
units of $\r_c=3M_p^2H^2/8\pi$, can thus be expressed as
\beq
{d(\r_g/\r_c)\over d\ln \om}=
 {dE_g\over \r_c a^3 d^3x d\ln \om}=\Om_G(\om, t)=
{1\over 3\pi ^2}
\left|\da_h(\om)\right|^2 \left(\om\over H\right)^2 .
\label{a16}
\eeq

\vskip 1 cm

\renewcommand{\theequation}{B.\arabic{equation}}
\setcounter{equation}{0}
\section{Appendix B \\ The spectral amplitude $S_h$}
\label{sec:B}
\noindent
Consider the Fourier transform of the polarized gravity wave
amplitude $h$,
\beq
h(\nu)=\int dt ~h(t) e^{-2\pi i\nu t}, ~~~~\nu=\om/2\pi . 
\label {b1}
\eeq
The average proper energy density, summing over polarizations,
can be written 
\bea
\r_g&=&{dE_g\over a^3 d^3x}={M_p^2\over 8\pi}
\langle \dot h^2\rangle=\nonumber\\
&=&-4\pi^2
{M_p^2\over  8\pi}\int d\nu d\nu'
\nu \nu' \langle h(\nu)h(\nu')\rangle
 e^{2\pi i (\nu+\nu')t} , 
\label{b2}
\eea
where the brackets denote time or ensemble average. By
defining the spectral amplitude $S_h(\nu)$, such that\cite{59}
\beq
\langle h(\nu) h^\ast (-\nu')\rangle= {1\over 2}
\da(\nu+\nu')S_h(\nu),
\label{b3}
\eeq
the energy distribution $\Om_g$, in critical units, can  be
expressed as
\beq
\Om_G(\nu,t)={1\over \r_c}{d\r_G\over d\ln \nu}=
{4\pi^2\over 3H^2(t)}\nu^3 S_h(\nu)
\label{b4}
\eeq
(I have assumed $S_h(\nu)=S_h(-\nu)$). 

A particular value of $S_h(\nu)$, representing the experimental
sensitivity at a given frequency $\nu$, corresponds to a present
graviton energy density\cite{32}
\beq
\Om_G(\nu, t_0)h_{100}^2\simeq 1,25\times 10^{36}\nu^3
S_h(\nu)~{\rm Hz}^{-2} .
\label{b5}
\eeq
Reaching the level of expected maximal  signal, $\Om_G h^2_{100}
~\laq~ 10^{-6}$ (see Sect. 4. 1), would thus require a 
sensitivity\cite{32} 
\beq
S_h^{1/2}(\nu)~\laq ~3 \times 10^{-26}\left({\rm
kHz}\over \nu \right)^{3/2}~{\rm Hz}^{-1/2}. 
\label{b6}
\eeq
Comparing eqs. (\ref{b4}) and (\ref{a16}), we can finally obtain a
useful relation between the spectral amplitude $S_h$ and the
dimensionless  amplitude $\da_h$,
\beq
S_h\left(\om\over 2\pi\right)={2\over
\pi\om}\left|\da_h(\om)\right|^2.
\label{b7}
\eeq

\end{document}